\documentclass[journal=jpccck,manuscript=article]{achemso}
\SectionNumbersOn

\usepackage[version=3]{mhchem} 
\usepackage{color}
\usepackage{graphics}
\usepackage{amsmath}
\usepackage{nccmath}
\usepackage{epsfig}
\usepackage{fancyhdr}
\usepackage{lineno}
\usepackage{float}
\usepackage[normalem]{ulem}
\usepackage{setspace}

\usepackage{wrapfig}  
\usepackage{amsmath,amssymb}
\usepackage{tabularx}

\usepackage{url} 
\usepackage{multicol}
\usepackage{multirow}
\usepackage[utf8]{inputenc}
\usepackage{graphicx}
\usepackage{lineno}

\author{Ahmed Gaber Abdelmagid}
\email{ahmed.abdelmagid@utu.fi}
\affiliation[Materials UTU]
{Department of Mechanical and Materials Engineering, University of Turku, FI-20014 Turku, Finland}

\author{Zhuoran Qiao}
\affiliation[Imperial chem]
{Department of Chemistry and Centre for Processable Electronics, Imperial College London, White City Campus, 82 Wood Lane, London, W12 0BZ UK}

\author{Boudewijn Coenegracht}
\affiliation[Materials UTU]
{Department of Mechanical and Materials Engineering, University of Turku, FI-20014 Turku, Finland}
\alsoaffiliation[]
{Current address: Molecular Materials and Nanosystems and Institute for Complex Molecular Systems, Eindhoven University of Technology}

\author{Gaon Yu}
\affiliation[Materials UTU]
{Department of Mechanical and Materials Engineering, University of Turku, FI-20014 Turku, Finland}
\alsoaffiliation[]
{Current address: Graduate Institute of Ferrous and Eco Materials Technology, Pohang University of Science and Technology, Pohang 37673, Republic of Korea}

\author{Hassan A. Qureshi}
\affiliation[Materials UTU]
{Department of Mechanical and Materials Engineering, University of Turku, FI-20014 Turku, Finland}

\author{Thomas D. Anthopoulos}
\affiliation[]
{Henry Royce Institute and Photon Science Institute, Department of Electrical and Electronic Engineering, The University of Manchester, Oxford Road, Manchester M13 9PL, UK}
\alsoaffiliation[]
{King Abdullah University of Science and Technology (KAUST), KAUST Solar Centre (KSC), Thuwal 23955-6900, Saudi Arabia}

\author{Nicola Gasparini}
\affiliation[Imperial chem]
{Department of Chemistry and Centre for Processable Electronics, Imperial College London, White City Campus, 82 Wood Lane, London, W12 0BZ UK}

\author{Konstantinos S. Daskalakis}
\email{konstantinos.daskalakis@utu.fi}
\affiliation[Materials UTU]
{Department of Mechanical and Materials Engineering, University of Turku, FI-20014 Turku, Finland}

\usepackage{graphicx} 

\title{Polaritons in non-fullerene acceptors for high responsivity angle-independent organic narrowband infrared  photodiodes}

\date{July 2024}

\begin{document}
\maketitle
\section{Abstract}
Narrowband infrared organic photodetectors are in great demand for sensing, imaging, and spectroscopy applications. However, most existing strategies for narrowband detection depend on spectral filtering either through saturable absorption, which requires active layers exceeding 500 nm, restricting the choice of materials for producing high-quality films, or cavity effects, which inherently introduce strong angular dispersion. Microcavity exciton-polariton (polariton) modes, which emerge from strong exciton-photon coupling, have recently been explored as an angular dispersion suppression strategy for organic optoelectronics. In this work, we present the first narrowband infrared polariton organic photodiode that combines angle-independent response with a record-high responsivity of 0.24 A/W at 965~nm and -2~V. 
This device, featuring a 100-nm-thin active layer comprising a non-fullerene acceptor, exhibits a detection mode with a full-width at half-maximum of less than 30 nm and a marginal angular dispersion of under 15 nm across $\pm$$45^\circ$. This study highlights the potential of polaritons as an innovative platform for developing next-generation optoelectronic devices that achieve simultaneous enhancements in optical and electronic performance.


\section{Main}

Photodiodes are essential solid-state devices that convert light into electrical signals, playing a pivotal role in photonics and materials research \cite{lan2022recent, zhang2023organic, li2024highly}. Their applications span from basic light detection to sophisticated multicolor imaging in telecommunications, machine vision, and microscopy, where color selectivity is a critical performance factor. Organic photodiodes (OPDs) have gained significant attention for their fabrication simplicity and mechanical flexibility \cite{wang2024semitransparent, wang2024flexible}. However, narrowband detection is challenging\cite{jansen2016organic, vanderspikken2021wavelength, wang2022narrowband, zhao2024narrowband}. Two primary strategies are employed to address this challenge. The first involves manipulating internal quantum efficiency by controlling charge collection efficiency (CCN) and exciton dissociation efficiency (EDN). This approach utilizes a thick bulk heterojunction (BHJ) layer to enable selective wavelength response by promoting recombination at shorter wavelengths \cite{armin2015narrowband, lin2015filterless, xie2020self, liu2022electron}.  The second strategy incorporates a microcavity (MC), which enhances absorption at specific wavelengths determined by the cavity's refractive index (n) and thickness (L)\cite{lupton2003organic, siegmund2017organic, tang2017polymer, wang2019organic, vanderspikken2022tuning, xing2021miniaturized, wang2021stacked, yang2021cavity}. These approaches have trade-offs: in the CCN approach, it is feasible with only a limited selection of organic semiconductor materials and often introduces challenges in processing, while microcavities inherit the OPD a strong angular dispersion \cite{siegmund2017organic, wang2021stacked, yang2021cavity}. However, compared to CCN, microcavities offer several distinct advantages making them a superior approach \cite{wang2022narrowband}. They can enhance spectral selectivity, extend responsivity, and achieve these enhancements within optimized OPD architectures that already exhibit excellent optoelectronic performance \cite{vanderspikken2021wavelength, wang2022narrowband, zhao2024narrowband}. Furthermore, their versatility allows for seamless integration into existing device designs\cite{Palo2023}.

An additional advantage of optical microcavities is their ability to facilitate strong coupling between excitons and cavity photons, forming hybrid states known as polaritons that inherit properties from light and matter \cite{frisk2019ultrastrong, garcia2021manipulating}. In OPDs, highly localized Frenkel excitons with a large effective mass are common. Reports indicate that under strong coupling, Frenkel excitons exhibit delocalization and low effective mass of cavity photons, leading to enhanced exciton diffusion \cite{coles2014polariton, xiang2020intermolecular,wang2021polariton, balasubrahmaniyam2023enhanced, sokolovskii2023multi, sandik2024cavity}, and suppression of exciton-exciton annihilation processes\cite{zhao2024stable, arneson2024color, qureshi2024giant}. These effects can enhance the electrical properties, stability, and overall functionality of organic optoelectronic devices. Moreover, polaritons can be viewed as cavity modes dressed with "heavy" excitons, which have a mass \( m_{\text{exc}} \). This coupling results in an angular dispersion that is significantly flatter than that of bare cavity photons with mass \( m_{\text{ph}} \). The reduced curvature reflects an increase in the effective mass of the polariton, \( m_{\text{pol}} \), which depends on the exciton-photon mixing ratio, as described by:

\[
\frac{1}{m_{\text{pol}}} = \frac{|C_{\text{ph}}|^2}{m_{\text{ph}}} + \frac{|C_{\text{exc}}|^2}{m_{\text{exc}}}
\]

\begin{figure}
\vspace{0pt}
\centering
\includegraphics[width=\linewidth]{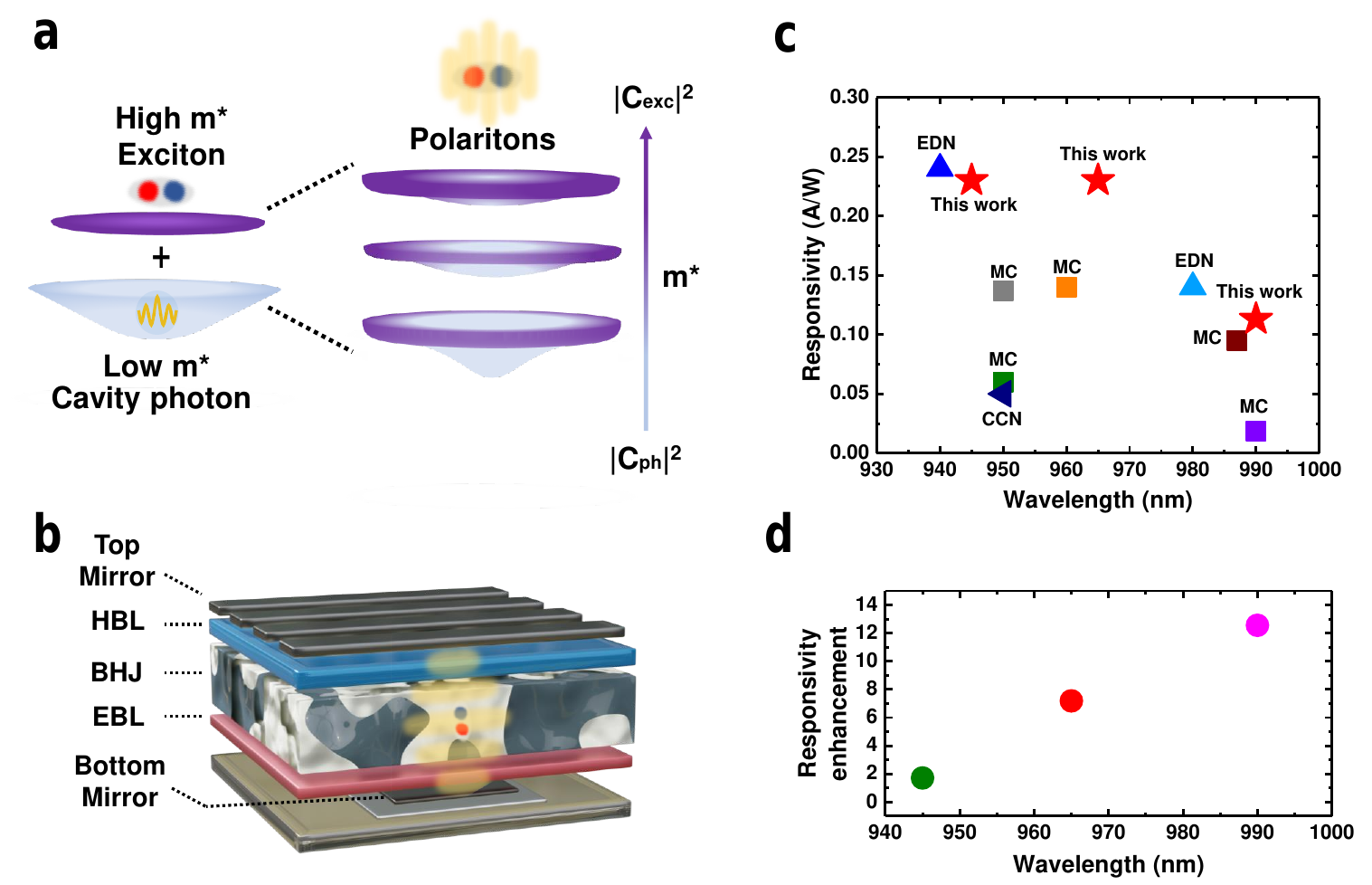}
\vspace{-12pt}
\caption{a) Schematic illustration of the impact of strong exciton-photon coupling on the effective mass and polariton dispersion. As the exciton content rises, the effective mass increases, resulting in a flatter dispersion. b) Simplified scheme of the polariton OPD with a sketch of the interaction between the cavity resonance and the exciton. c) Comparison of the responsivities of narrowband OPDs at 0 V with peak wavelengths between 940 nm and 990 nm (Supplementary Table~\ref{tab:responsivity}). d) The responsivity enhancement of the polariton OPDs with device resonance at 945~nm (olive), 965~nm (red), 990~nm (magenta). The developed polaritonic OPD constructed from layers of (glass/ITO (Electrode)/Ag 25 nm (Bottom Mirror)/MoO$_3$ 10 nm (EBL)/PTB7-Th:IEICO-4F (BHJ)/C$_{60}$:LiF 25 nm (HBL)/Ag 150 nm (Top Mirror/Electrode)), with light entering from the ITO side.}
\label{fig:1}
\vspace{0pt}
\end{figure}

Here, \( |C_{\text{ph}}|^2 \) and \( |C_{\text{exc}}|^2 \) are the Hopfield coefficients that denote the photon and exciton fractions of the polariton mode, respectively, with \( |C_{\text{ph}}|^2 + |C_{\text{exc}}|^2 = 1 \). By increasing the exciton content \( |C_{\text{exc}}|^2 \), the effective mass \( m_{\text{pol}} \) increases, leading to a suppression of the polariton dispersion (Fig.~\ref{fig:1}a) \cite{deng2010exciton}. Increasing the exciton content in the polariton mode can be achieved by either increasing the cavity-exciton resonance overlap (detuning) or the coupling strength of the system \cite{deng2010exciton, frisk2019ultrastrong}. This approach has recently been used to suppress angular dispersion in infrared polariton microcavities \cite{eizner2018organic}, carbon nanotube photodiodes \cite{mischok2020spectroscopic}, optical filters\cite{mischok2024breaking}, and organic light-emitting diodes \cite{mischok2023highly, de2024organic}. However, a significant challenge remains: demonstrating polaritonic optoelectronic devices that not only outperform their non-polaritonic counterparts but also exceed the performance benchmarks typically reported in the field. Achieving such breakthrough results is crucial for realizing the full potential of polariton-based technologies in practical applications.

Here, we demonstrate, for-the-first-time, a narrowband polariton OPD (Fig.~\ref{fig:1}b) that utilizes the polariton photonic gains while reaching a high responsivity in the infrared region from $\sim$ 945 - 990~nm (Fig.~\ref{fig:1}c). Note, here we consider as narrowband a detection bandwidth with a full-width at half-maximum (FWHM) of less than 100~nm\cite{zhao2024narrowband}, and in our reported devices this was less than 45~nm. In particular, at 965~nm we recorded responsivity of {0.23 A/W} at 0~V, which is the highest reported to date \cite{xie2020self, armin2015narrowband, xing2021miniaturized, siegmund2017organic, tang2017polymer, vanderspikken2022tuning, wang2021stacked, liu2022electron} (see Fig.~\ref{fig:1}c for devices operating at 945 and 990 nm). Importantly, we achieved these metrics with 100-nm active layer OPDs which are substantially thinner than prior state-of-the-art (see Supplementary Table~\ref{tab:responsivity}). Interestingly, compared with the reference device (see Methods), these represent a 7-fold enhancement at 965 nm (see Fig.~\ref{fig:1}d and Supplementary Fig.~\ref{Ref vs Polariton OPDs}). Importantly, these polariton OPDs achieved such optical response while maintaining low angular dispersion of less than 17 nm within an observation cone of $\pm$$45^\circ$( see analysis in Section 3.2 and Fig.~\ref{fig:3}).

In this work, the active medium of the polariton OPDs is a blend of PTB7-Th:IEICO-4F. We selected IEICO-4F because of its high oscillator strength \cite{souza2022binding}, and good film morphology when blended with PTB7-Th\cite{song2018controlling} - both essential for reaching strong coupling in planar microcavities \cite{bhuyan2023rise}. IEICO-4F belongs to the class of nonfullerene acceptors (NFAs) that offer future possibilities to tune their absorption resonance down to the visible wavelength range via molecular design \cite{saleem2021designing, zhong2024general, wadsworth2019critical}. Furthermore, NFAs have been extensively exploited in organic optoelectronic devices, enabling efficient organic photovoltaics (OPVs) \cite{moustafa2023thermal, zhu2022single, chen202319, sun2024pi} and broadband OPDs \cite{yang2021mitigating, siddik2023interface, park2023state, luong2024highly} due to their high charge mobility, efficient charge transport, and chemical and thermal stability.

\begin{figure}[h!]
\vspace{0pt}
\centering
\includegraphics[width=\linewidth]{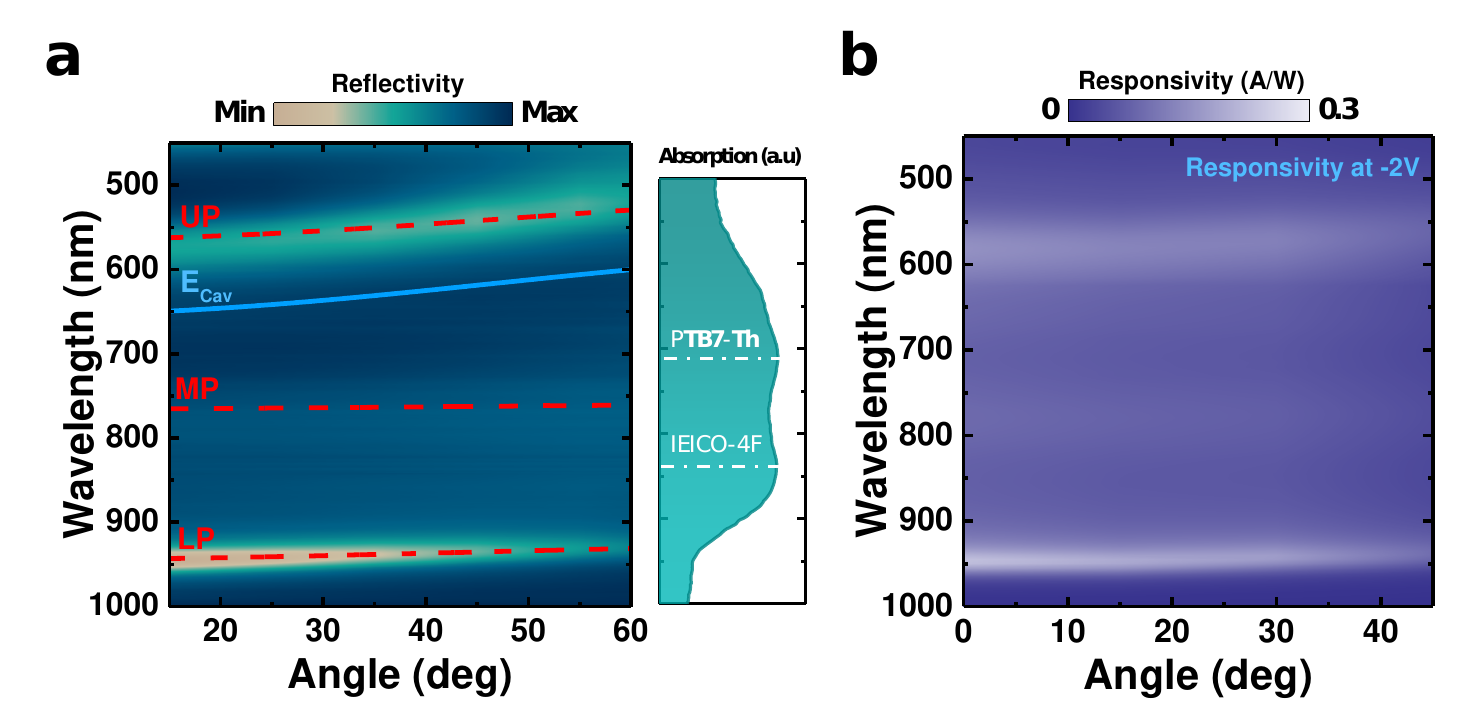}
\vspace{-12pt}
\caption{ a) Polariton characteristics of the polariton OPD. Angle-resolved reflectivity of polariton OPD with device resonance of 945~nm from $15^\circ$ to $60^\circ$. The solid blue line is the cavity energy dispersion, and the dashed red lines are fitted polariton dispersions. Besides is the absorption of the active blend and the dashed white line is the molecular exciton energies of PTB7-Th and IEICO-4F. c) Polariton characteristics under negative electrical bias. The angle-resolved responsivity of the polariton device from $0^\circ$ to $45^\circ$ at -2~V.}
\label{SC POPD}
\vspace{0pt}
\end{figure}

\begin{figure}[h!]
\vspace{0pt}
\centering
\includegraphics[width=\linewidth]{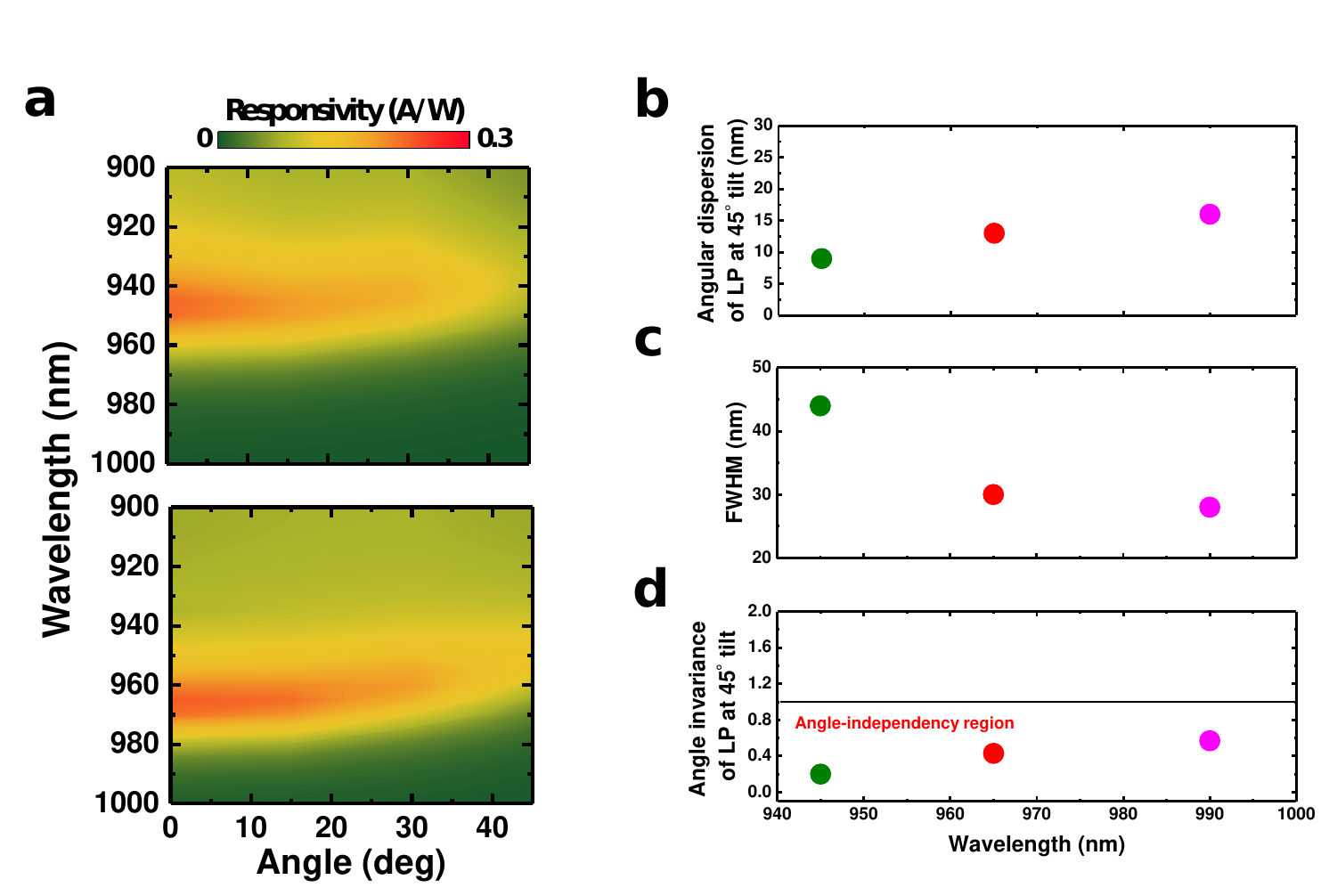}
\vspace{-12pt}
\caption{a) Angle-resolved responsivity maps from $0^\circ$ to $45^\circ$ with an interval of $15^\circ$ for the polariton OPDs with device resonance of 945 nm (top) and 965 nm (bottom) b) Angular dispersion,  c) FWHM, and d) Angle invariance for the polariton OPDs with device resonance of 920~nm (olive), 965~nm (red), and 990~nm (magenta).}
\label{fig:3}
\vspace{0pt}
\end{figure}

Figure~\ref{SC POPD}a presents the angle-resolved reflectivity map of a polariton OPD with an active layer thickness of $\sim$92~nm (see Methods). There, three distinct reflectivity dips appear, corresponding to the upper, middle, and lower polariton branches, respectively. In this study, we focus on the lower polariton mode, as its resonance falls within the investigated spectral range. Henceforth, "device resonance" refers to the lower polariton resonance. The corresponding angle-resolved reflectivity spectra are shown in Supplementary Fig.~\ref{POPD ref spectra}. The presence of three polariton branches indicates effective coupling between the cavity mode and the molecular excitons of PTB7-Th (710~nm) and IEICO-4F (835~nm), as highlighted by the dashed white lines. By fitting the data with a coupled harmonic oscillator (CHO) model, we determined the cavity resonance at 656~nm at an incidence angle of 0°; the cavity dispersion is shown as a solid blue line. The fitted polariton dispersions are depicted as dashed red lines. For these modes, our fitting result to a Rabi splitting of 0.57~eV and 0.31~eV and detuning of $\Delta$ = $E_{c}$ - $E_{IEICO-4F}$ = 0.41 eV. Supplementary Fig.~\ref{POPD945 fraction} illustrates the exciton and photon fractions of the lower polariton (LP) branch, revealing a significant exciton contribution from IEICO-4F, approximately 60\%. This high exciton content enhances the effective mass of the polariton and results in an exciton-like optical response, as seen in Fig.~\ref{SC POPD}a \cite{deng2010exciton}. Additional analysis for polariton OPDs with device resonances at 965~nm and 990~nm are provided in the Supplementary Information.

To verify that the device operates in the strong coupling regime under bias, we performed angle-resolved responsivity measurements for the polariton OPD operating at 945~nm at -2~V. As shown in Fig.~\ref{SC POPD}b, there is excellent agreement between reflectivity and responsivity measurements. Prior to device fabrication, we optimized polariton microcavities using neat films of IEICO-4F and Y6. These results are shown in Supplementary Fig.~\ref{MC NFAs}. We selected IEICO-4F due to its well-documented performance in the literature\cite{yang2021mitigating, siddik2023interface}.

Figure~\ref{fig:3}a shows angle-resolved responsivity maps from polariton OPDs with device resonance of 945~nm (top) and 965~nm (bottom). Our devices exhibit responsivities above 0.096 A/W within their acceptance cone of $\pm$$45^\circ$. Within this range, the responsivity peak shifts by 9 nm for the thinnest OPDs and up to 16 nm for the thickest (Fig.~\ref{fig:3}b), which is substantially suppressed compared to reported narrowband conventional microcavity-enhanced OPDs \cite{siegmund2017organic, wang2021stacked, yang2021cavity}. This reduced angular dependency highlights the benefits of strong coupling within our devices and is directly correlated with the exciton content in the LP (see Supplementary Fig.~\ref{LP ex fraction}). Because angle-resolved responsivity measurement was limited to 5~nm spectral resolution, it overestimated the angle dispersion of our devices. To circumvent this, we used an in-house built angle-resolved reflectivity setup with a spectral resolution of 0.4~nm \cite{abdelmagid2024identifying}. 

At normal incidence, the FWHM of the responsivity peak is less than 45 nm for all devices (Fig.~\ref{fig:3}c). This narrow FWHM is achieved by tuning the mode to the above-gap, low-absorption region of the active material, where reduced absorption allows for enhanced narrowband polariton effects with strong angular stability. We highlight this effect in Fig.~\ref{fig:3}d where we plot the angle invariance factor, defined as angular dispersion over the FWHM, with wavelength, showing that all devices are in principle angle independent. 

We explored the limitations of our device architecture by tuning the resonance to shorter wavelengths at 920~nm. This resulted in a significantly broader FWHM of more than 100 nm and a reduced responsivity compared to the reference device (Supplementary Fig.~\ref{Polariton OPD920}). We speculate that this is because the mode in this device overlaps substantially with a high-absorption region of the active material, which dampens the resonance and reduces both the polariton-enhanced spectral narrowing and the overall device efficiency. \cite{lupton2003organic, lin2015filterless, siegmund2017organic}.

\begin{figure}[h!]
\vspace{0pt}
\centering
\includegraphics[width=\linewidth]{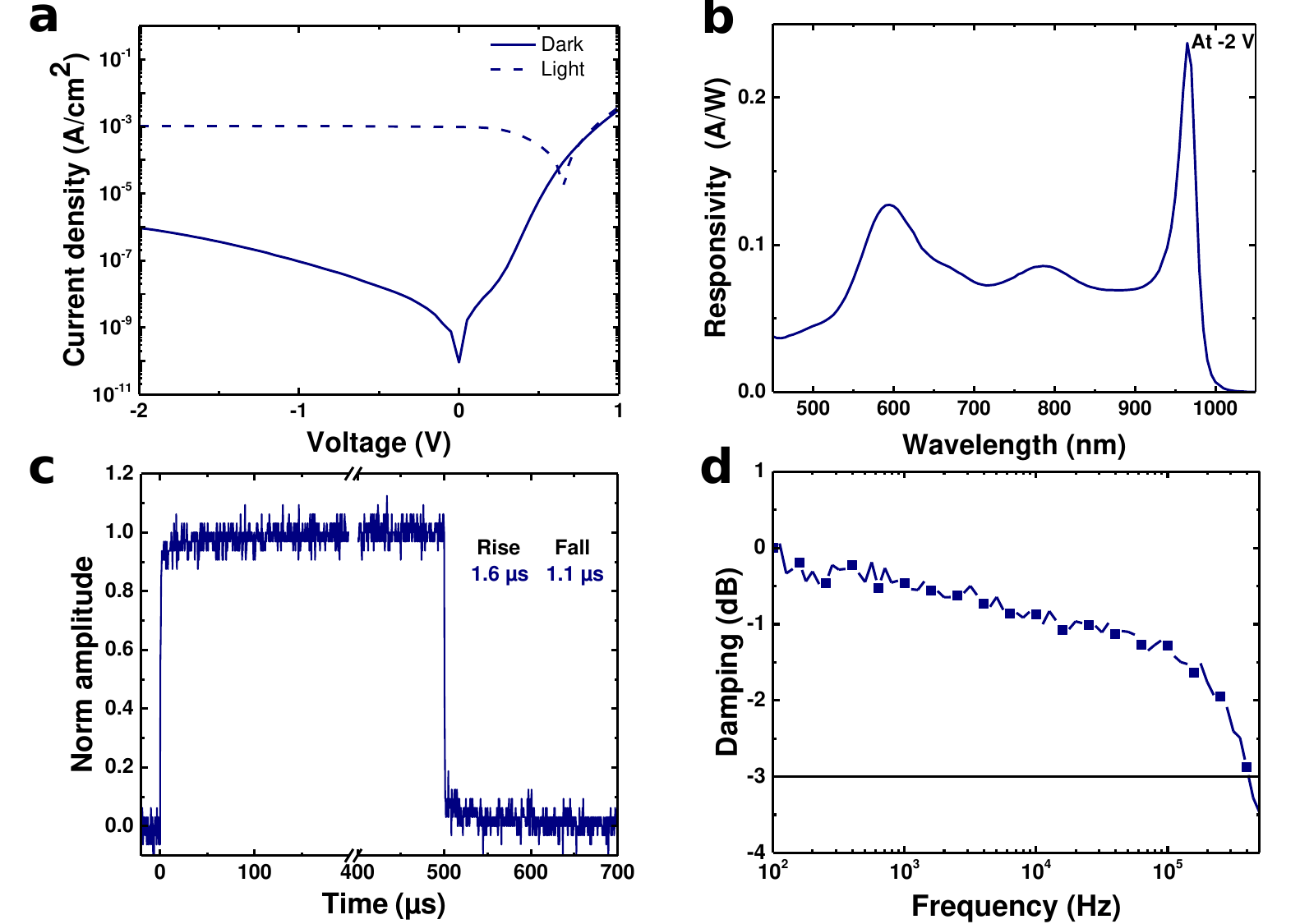}
\vspace{-12pt}
\caption{a) Dark and light J-V curve of the optimized device, polariton OPD with resonance wavelength at 965~nm. b) Responsivity at -2~V. c) Transient photoresponse of the polariton OPD at -2~V under 970 nm NIR light. d) Cut-off
frequency of the polariton OPD at -3~dB at -2~V under 970 nm NIR light.} 
\label{fig:4}
\vspace{0pt}
\end{figure}

Figure~\ref{fig:4}a shows the characteristic current-voltage measurement from polariton OPD with resonance wavelength 965~nm in dark and illuminated conditions. A dark current density of $9.29\times10^{-7}$ A $cm^{-2}$ at -2~V was recorded. This relatively high value is attributed to the lack of optimization of the hole and electron blocking layers, which is beyond the scope of this study. Under the illumination conditions (see Methods), we observe a photocurrent density (J$_{L}$) value of $1.04\times10^{-3}$ A $cm^{-2}$ at -2~V. 

As previously discussed, we have achieved a maximum responsivity of 0.23 A/W at 965 nm and 0~V, which peaked at 0.24 A/W when biased at -2~V, which is the typical operating conditions of OPDs \cite{hu2024remarkable}. Fig~\ref{fig:4}b presents the responsivity spectra of the polariton OPD  at -2~V, while Supplementary Fig~\ref{POPD Rs spectra} displays the responsivity across various bias voltages. Such bias independence indicates an enhanced charge extraction and collection within the active layer under reverse bias. \cite{du2024high}.

The specific detectivity ($D^{*}$) can be calculated using the responsivity and noise spectral density (see Methods section). However, $D^{*}$ can be overestimated when shot noise, which is derived from dark current, is considered solely as the origin of noise since other sources of noise including flicker and thermal noise, will also contribute significantly to the overall noise in the device \cite{jacoutot2023enhanced}. Therefore, in this work, noise spectral density was calculated through the fast Fourier transform (FFT) of the dark current (Supplementary Fig~\ref{Polariton OPD noise}). We calculated $D^{*}_{\text{measured}}$ to be $5.93\times10^{9}$ Jones at -2~V at 965~nm. 


Dynamic measurements were conducted to assess the operation speed of the polariton OPD at -2~V. As shown in Figure 5c, The polariton OPD exhibited a rise time of 1.6 µs and a fall time of 1.1 µs under IR light illumination (Fig~\ref{fig:4}c). Additionally, the cut-off frequency, defined as the frequency at which the photocurrent response diminishes by 3 dB from its low-frequency value, was measured to be 412 kHz under 970 nm illumination (Fig~\ref{fig:4}d). These results demonstrate the potential of the polariton OPD to be applied for fast imaging applications \cite{jansen2016organic, jacoutot2023enhanced}.

For comparison, the reference device exhibited a broadband response with spectral responsivity exceeding 0.40 A/W and an EQE of over 65\% in the 760–860 nm range. However, these values declined beyond 900 nm due to the reduced absorption of IEICO-4F. The responsivity of the reference device, measured under different bias, is shown in Supplementary Fig~\ref{ref Rs spectra}. In contrast, as previously demonstrated, the responsivity spectra of the polariton OPD differed significantly from those of the reference device. Despite this, all key performance metrics of the polariton OPD were comparable to those of the reference device (see Supplementary Fig~\ref{comparison}), indicating that the integration of polaritons into the broadband device did not negatively impact its performance. 


\newpage
\section{Conclusions and outlook}

In this work, we have demonstrated the first narrowband polariton organic photodiodes with responsivity and external quantum efficiency (EQE) that surpass the performance of previously reported narrowband OPDs in the near-infrared (NIR) region. By employing a Fabry-Pérot cavity design and leveraging the strong coupling regime, we achieved narrowband detection with FWHM of less than 45 nm and responsivity values up to 0.24 A/W at 965 nm, at -2~V bias. The devices showed minimal dispersion across a wide angular acceptance cone of $\pm45^\circ$, showcasing the angle-independent nature of our devices.

Our work highlights the potential of integrating polaritonic enhancements in OPDs, demonstrating their ability to improve spectral selectivity and suppress angular dispersion without compromising their figures of merit. The choice of non-fullerene acceptor IEICO-4F was instrumental in achieving strong coupling in the NIR spectrum because of its high oscillator strength in this spectral range. Additionally, our findings confirm that polaritonic devices can maintain performance metrics comparable to, or even exceeding, those of conventional broadband OPDs. Moreover, NFAs provide a versatile platform for expanding the operating wavelength range, as their absorption properties can be molecularly tailored to cover both the visible and infrared regions \cite{saleem2021designing, zhong2024general}. This study demonstrates that OPDs represent an exceptionally promising device configuration where polaritonic effects can lead to substantial and undisputed improvements in performance. Our approach lays the groundwork for the development of high-performance, angle-independent polaritonic optoelectronic devices, offering promising applications in imaging, sensing, and spectroscopy.

\nolinenumbers


\newpage
\section{Methods}
\subsection{Materials}
PTB7-Th, IEICO-4F, and Y6 were purchased from 1-material. $C_{60}$ was purchased from Lumtec.
\subsection{Microcavity and device fabrication}
Glass substrates prepatterned with ITO (15mm*15mm*0.7mm) were cleaned with water/soap (3 \% Decon 90), acetone, and Isopropanol solutions to remove any residue from the surfaces. The substrates were sonicated for 10 minutes on each step and finally dried with a $N_{2}$ purge. A similar cleaning protocol was deployed for the silicon substrates that were used to measure the thicknesses with ellipsometry. 
\newline\textbf{Thin-film microcavities:} IEICO-4F and Y6 were dissolved in chloroform at a concentration of 16 mg/ml, then stirred for two hours at 80°C, followed by one hour at 60°C inside a glovebox. Thin-film microcavities were fabricated by spin-coating the NFA solution onto a 120 nm Ag layer at 2500 rpm for Y6 and 3000 rpm for IEICO-4F, each for 30 seconds. The films were then annealed at 100°C for 10 minutes in the glovebox, followed by the evaporation of a 30 nm Ag layer as the top mirror.
\newline \textbf{Device fabrication:} polariton OPDs were fabricated in the architecture of (glass/ITO/Ag 25 nm/MoO$_{3}$ 10 nm/PTB7-Th:IEICO-4F/$C_{60}$:LiF 25 nm/Ag 150 nm). The bottom silver mirror and MoO$_{3}$ were deposited using an Angstrom evaporation system. The photoactive blend (PTB7-Th:IEICO-4F) was deposited in a glovebox. the blend was spin-coated from a 25 mg/ml solution in chlorobenzene:chloronaphthaline (96:4 v/v)at different rpm with the same annealing protocol used for the neat NFA films. The active layer thicknesses were varied by varying the rpm. On top of the photoactive layer, a $C_{60}$:LiF (1:4) layer was then deposited via co-evaporation. Finally, 150 nm-thick Ag was deposited as the top electrode by evaporation through a shadow mask giving photodiodes with pixel areas of 0.08 $cm^{2}$. The reference device was identical to the polariton device but did not include the thin Ag layer. All devices were encapsulated using a glass cover-slip and a UV-curable epoxy (Ossila Ltd) prior to characterization.

\subsection{Optical Characterization}
The thicknesses of the thin films were acquired using a J.A Wollam VASE ellipsometer. We utilized an Xe lamp with a spectral range of 250-2500 nm to obtain the spectra, the data was analyzed by fitting a Cauchy model in the transparent region of the film. The dispersion of the microcavities and devices was obtained with the VASE ellipsometer and a custom-built angle-resolved imaging setup. The setup uses a collimated light from a halogen lamp illuminating the sample through a 0.75 NA microscope objective. The reflected light is then collected with the same objective and the back focal plane image is then focused into the slit of the spectrometer that is coupled to a two-dimensional (2D) CCD camera (1340*400 pixels). The reflectivity dispersion was then resolved in wavelength vs angle. The couple harmonic oscillator model was used to fit the polariton modes and extract the Rabi splitting energy, cavity resonance, and Hoppfield coefficients.

\subsection{Electrical Characterization}
J-V measurements were conducted using a Keithley 4200 Source-Measure unit (scan rate 50 mV s–1).  An Oriel Instruments Solar Simulator with a Xenon lamp and calibrated to a silicon reference cell was used to provide AM1.5G irradiance. 
External quantum efficiency (EQE) was measured using an integrated system from Quantum Design PV300 with a modulation frequency of 90 Hz. Responsivity was further calculated using equation
\begin{equation}
R = \text{EQE} \cdot \frac{q \lambda}{h c}
\end{equation}
Where q is elementary charge, $\lambda$ is the wavelength of the light, h is Planck constant, and c is the speed of light. All the devices were tested in ambient air.
Specific detectivity was calculated using equation
\begin{equation}
D^* = \frac{R \sqrt{A \Delta f}}{i_n}
\end{equation}
Where R is responsivity, A denoted the device photoactive area, $\Delta$f is the detection bandwidth and $i_{n}$ is the noise spectral density. Here, to accurately evaluate specific detectivity, $i_{n}$  was calculated by the fast Fourier transform of the dark current obtained from a digital oscilloscope (Siglent, SDS6054A) with combination of a FEMTO-100 preamplifier under -2 V. The specific detectivity value can be overestimated by considering solely the contribution of dark current ($i_{shot}$) to the noise using equation
\begin{equation}
i_{shot}=\sqrt{2qI_{d}}
\end{equation}
Where $I_{d}$ refers to the dark current.
Dynamic measurements were performed using the digital oscilloscope (Siglent, SDS6054A). The OPDs were illuminated with a 970 nm LED driven by a function generator (ThorLabs DC2200). For determination of the rise and fall time, which is defined as the time interval in which the output photocurrent increases from 10\% to 90\% and diminishes from 90\% down to 10\% of the peak photocurrent, a 1 kHz square wave pulse was applied to the LED using the function generator. For determination of the cut-off frequency, sinusoidal functions with varying frequencies between 100 Hz and 1 MHz were used to drive the LED from the function generator. All the devices were tested in ambient air.

\subsection*{Author Contributions}
AGA, TDA, and KSD initiated the project. AGA designed the project, fabricated the devices, performed the majority of experiments, and analyzed the data. ZQ together with AGA performed the optoelectronic measurements, supervised by NG. HAQ performed the high-resolution reflectivity measurements. BC and GY contributed to the sample fabrication and characterization. AGA and KSD wrote the manuscript. KSD supervised the project. All authors contributed to the draft, discussion, and analysis of the data.

\subsection*{Conflicts of interest}
There are no conflicts to declare.

\begin{acknowledgement}
This project has received funding from the European Research Council (ERC) under the European Union’s Horizon 2020 research and innovation programme (grant agreement No. [948260]) and by the European Innovation Council through the project SCOLED (Grant Agreement Number 101098813). Views and opinions expressed are however those of the author(s) only and do not necessarily reflect those of the European Union or the European Innovation Council and SMEs Executive Agency (EISMEA). Neither the European Union nor the granting authority can be held responsible for them. ZQ and NG thank the King Abdullah University of Science and Technology (KAUST) Office of Sponsored Research (OSR) under Award No. OSR-2020-CRG8-4095 and ORFS-2023-OFP-5544.
\end{acknowledgement}

\bibliography{references}  

\providecommand{\latin}[1]{#1}
\makeatletter
\providecommand{\doi}
  {\begingroup\let\do\@makeother\dospecials
  \catcode`\{=1 \catcode`\}=2 \doi@aux}
\providecommand{\doi@aux}[1]{\endgroup\texttt{#1}}
\makeatother
\providecommand*\mcitethebibliography{\thebibliography}
\csname @ifundefined\endcsname{endmcitethebibliography}  {\let\endmcitethebibliography\endthebibliography}{}
\begin{mcitethebibliography}{58}
\providecommand*\natexlab[1]{#1}
\providecommand*\mciteSetBstSublistMode[1]{}
\providecommand*\mciteSetBstMaxWidthForm[2]{}
\providecommand*\mciteBstWouldAddEndPuncttrue
  {\def\EndOfBibitem{\unskip.}}
\providecommand*\mciteBstWouldAddEndPunctfalse
  {\let\EndOfBibitem\relax}
\providecommand*\mciteSetBstMidEndSepPunct[3]{}
\providecommand*\mciteSetBstSublistLabelBeginEnd[3]{}
\providecommand*\EndOfBibitem{}
\mciteSetBstSublistMode{f}
\mciteSetBstMaxWidthForm{subitem}{(\alph{mcitesubitemcount})}
\mciteSetBstSublistLabelBeginEnd
  {\mcitemaxwidthsubitemform\space}
  {\relax}
  {\relax}

\bibitem[Lan \latin{et~al.}(2022)Lan, Lee, and Zhu]{lan2022recent}
Lan,~Z.; Lee,~M.-H.; Zhu,~F. Recent advances in solution-processable organic photodetectors and applications in flexible electronics. \emph{Advanced Intelligent Systems} \textbf{2022}, \emph{4}, 2100167\relax
\mciteBstWouldAddEndPuncttrue
\mciteSetBstMidEndSepPunct{\mcitedefaultmidpunct}
{\mcitedefaultendpunct}{\mcitedefaultseppunct}\relax
\EndOfBibitem
\bibitem[Zhang \latin{et~al.}(2023)Zhang, Jiang, Feng, Song, and Shen]{zhang2023organic}
Zhang,~X.; Jiang,~J.; Feng,~B.; Song,~H.; Shen,~L. Organic photodetectors: materials, device, and challenges. \emph{Journal of Materials Chemistry C} \textbf{2023}, \emph{11}, 12453--12465\relax
\mciteBstWouldAddEndPuncttrue
\mciteSetBstMidEndSepPunct{\mcitedefaultmidpunct}
{\mcitedefaultendpunct}{\mcitedefaultseppunct}\relax
\EndOfBibitem
\bibitem[Li \latin{et~al.}(2024)Li, Hu, Wu, Ding, Zhang, Sun, Li, Liu, Shao, Fang, \latin{et~al.} others]{li2024highly}
Li,~T.; Hu,~G.; Wu,~H.; Ding,~L.; Zhang,~J.; Sun,~M.; Li,~Y.; Liu,~Z.; Shao,~Y.; Fang,~Y. \latin{et~al.}  Highly sensitive water pollution monitoring using colloid-processed organic photodetectors. \emph{Nature Water} \textbf{2024}, 1--12\relax
\mciteBstWouldAddEndPuncttrue
\mciteSetBstMidEndSepPunct{\mcitedefaultmidpunct}
{\mcitedefaultendpunct}{\mcitedefaultseppunct}\relax
\EndOfBibitem
\bibitem[Wang \latin{et~al.}(2024)Wang, Zhang, Samigullina, Winkler, Dollinger, Kublitski, Jia, Ji, Reineke, Spoltore, \latin{et~al.} others]{wang2024semitransparent}
Wang,~Y.; Zhang,~T.; Samigullina,~D.; Winkler,~L.~C.; Dollinger,~F.; Kublitski,~J.; Jia,~X.; Ji,~R.; Reineke,~S.; Spoltore,~D. \latin{et~al.}  Semitransparent Near-Infrared Organic Photodetectors: Flexible, Large-Area, and Physical-Vapor-Deposited for Versatile Advanced Optical Applications. \emph{Advanced Functional Materials} \textbf{2024}, 2313689\relax
\mciteBstWouldAddEndPuncttrue
\mciteSetBstMidEndSepPunct{\mcitedefaultmidpunct}
{\mcitedefaultendpunct}{\mcitedefaultseppunct}\relax
\EndOfBibitem
\bibitem[Wang \latin{et~al.}(2024)Wang, Cheng, Fukuda, Hu, Xu, and Someya]{wang2024flexible}
Wang,~Z.; Cheng,~S.; Fukuda,~K.; Hu,~W.; Xu,~X.; Someya,~T. Flexible Near-Infrared Organic Photodetectors for Emergent Wearable Applications. \emph{Wearable Electronics} \textbf{2024}, \relax
\mciteBstWouldAddEndPunctfalse
\mciteSetBstMidEndSepPunct{\mcitedefaultmidpunct}
{}{\mcitedefaultseppunct}\relax
\EndOfBibitem
\bibitem[Jansen-van Vuuren \latin{et~al.}(2016)Jansen-van Vuuren, Armin, Pandey, Burn, and Meredith]{jansen2016organic}
Jansen-van Vuuren,~R.~D.; Armin,~A.; Pandey,~A.~K.; Burn,~P.~L.; Meredith,~P. Organic photodiodes: the future of full color detection and image sensing. \emph{Advanced Materials} \textbf{2016}, \emph{28}, 4766--4802\relax
\mciteBstWouldAddEndPuncttrue
\mciteSetBstMidEndSepPunct{\mcitedefaultmidpunct}
{\mcitedefaultendpunct}{\mcitedefaultseppunct}\relax
\EndOfBibitem
\bibitem[Vanderspikken \latin{et~al.}(2021)Vanderspikken, Maes, and Vandewal]{vanderspikken2021wavelength}
Vanderspikken,~J.; Maes,~W.; Vandewal,~K. Wavelength-selective organic photodetectors. \emph{Advanced Functional Materials} \textbf{2021}, \emph{31}, 2104060\relax
\mciteBstWouldAddEndPuncttrue
\mciteSetBstMidEndSepPunct{\mcitedefaultmidpunct}
{\mcitedefaultendpunct}{\mcitedefaultseppunct}\relax
\EndOfBibitem
\bibitem[Wang \latin{et~al.}(2022)Wang, Kublitski, Xing, Dollinger, Spoltore, Benduhn, and Leo]{wang2022narrowband}
Wang,~Y.; Kublitski,~J.; Xing,~S.; Dollinger,~F.; Spoltore,~D.; Benduhn,~J.; Leo,~K. Narrowband organic photodetectors--towards miniaturized, spectroscopic sensing. \emph{Materials Horizons} \textbf{2022}, \emph{9}, 220--251\relax
\mciteBstWouldAddEndPuncttrue
\mciteSetBstMidEndSepPunct{\mcitedefaultmidpunct}
{\mcitedefaultendpunct}{\mcitedefaultseppunct}\relax
\EndOfBibitem
\bibitem[Zhao \latin{et~al.}(2024)Zhao, Wang, Liu, Ma, and Zhang]{zhao2024narrowband}
Zhao,~X.; Wang,~J.; Liu,~M.; Ma,~X.; Zhang,~F. Narrowband Organic Photodetectors: From Fundamentals to Prospects. \emph{Advanced Optical Materials} \textbf{2024}, 2401087\relax
\mciteBstWouldAddEndPuncttrue
\mciteSetBstMidEndSepPunct{\mcitedefaultmidpunct}
{\mcitedefaultendpunct}{\mcitedefaultseppunct}\relax
\EndOfBibitem
\bibitem[Armin \latin{et~al.}(2015)Armin, Jansen-van Vuuren, Kopidakis, Burn, and Meredith]{armin2015narrowband}
Armin,~A.; Jansen-van Vuuren,~R.~D.; Kopidakis,~N.; Burn,~P.~L.; Meredith,~P. Narrowband light detection via internal quantum efficiency manipulation of organic photodiodes. \emph{Nature communications} \textbf{2015}, \emph{6}, 6343\relax
\mciteBstWouldAddEndPuncttrue
\mciteSetBstMidEndSepPunct{\mcitedefaultmidpunct}
{\mcitedefaultendpunct}{\mcitedefaultseppunct}\relax
\EndOfBibitem
\bibitem[Lin \latin{et~al.}(2015)Lin, Armin, Burn, and Meredith]{lin2015filterless}
Lin,~Q.; Armin,~A.; Burn,~P.~L.; Meredith,~P. Filterless narrowband visible photodetectors. \emph{Nature Photonics} \textbf{2015}, \emph{9}, 687--694\relax
\mciteBstWouldAddEndPuncttrue
\mciteSetBstMidEndSepPunct{\mcitedefaultmidpunct}
{\mcitedefaultendpunct}{\mcitedefaultseppunct}\relax
\EndOfBibitem
\bibitem[Xie \latin{et~al.}(2020)Xie, Xie, Zhang, Yin, Hu, Yu, Huang, and Cao]{xie2020self}
Xie,~B.; Xie,~R.; Zhang,~K.; Yin,~Q.; Hu,~Z.; Yu,~G.; Huang,~F.; Cao,~Y. Self-filtering narrowband high performance organic photodetectors enabled by manipulating localized Frenkel exciton dissociation. \emph{Nature communications} \textbf{2020}, \emph{11}, 2871\relax
\mciteBstWouldAddEndPuncttrue
\mciteSetBstMidEndSepPunct{\mcitedefaultmidpunct}
{\mcitedefaultendpunct}{\mcitedefaultseppunct}\relax
\EndOfBibitem
\bibitem[Liu \latin{et~al.}(2022)Liu, Zeiske, Jiang, Desta, Mertens, Gielen, Shanivarasanthe, Boyen, Armin, and Vandewal]{liu2022electron}
Liu,~Q.; Zeiske,~S.; Jiang,~X.; Desta,~D.; Mertens,~S.; Gielen,~S.; Shanivarasanthe,~R.; Boyen,~H.-G.; Armin,~A.; Vandewal,~K. Electron-donating amine-interlayer induced n-type doping of polymer: nonfullerene blends for efficient narrowband near-infrared photo-detection. \emph{Nature Communications} \textbf{2022}, \emph{13}, 5194\relax
\mciteBstWouldAddEndPuncttrue
\mciteSetBstMidEndSepPunct{\mcitedefaultmidpunct}
{\mcitedefaultendpunct}{\mcitedefaultseppunct}\relax
\EndOfBibitem
\bibitem[Lupton \latin{et~al.}(2003)Lupton, Koeppe, M{\"u}ller, Feldmann, Scherf, and Lemmer]{lupton2003organic}
Lupton,~J.~M.; Koeppe,~R.; M{\"u}ller,~J.~G.; Feldmann,~J.; Scherf,~U.; Lemmer,~U. Organic microcavity photodiodes. \emph{Advanced Materials} \textbf{2003}, \emph{15}, 1471--1474\relax
\mciteBstWouldAddEndPuncttrue
\mciteSetBstMidEndSepPunct{\mcitedefaultmidpunct}
{\mcitedefaultendpunct}{\mcitedefaultseppunct}\relax
\EndOfBibitem
\bibitem[Siegmund \latin{et~al.}(2017)Siegmund, Mischok, Benduhn, Zeika, Ullbrich, Nehm, B{\"o}hm, Spoltore, Fr{\"o}b, K{\"o}rner, \latin{et~al.} others]{siegmund2017organic}
Siegmund,~B.; Mischok,~A.; Benduhn,~J.; Zeika,~O.; Ullbrich,~S.; Nehm,~F.; B{\"o}hm,~M.; Spoltore,~D.; Fr{\"o}b,~H.; K{\"o}rner,~C. \latin{et~al.}  Organic narrowband near-infrared photodetectors based on intermolecular charge-transfer absorption. \emph{Nature communications} \textbf{2017}, \emph{8}, 15421\relax
\mciteBstWouldAddEndPuncttrue
\mciteSetBstMidEndSepPunct{\mcitedefaultmidpunct}
{\mcitedefaultendpunct}{\mcitedefaultseppunct}\relax
\EndOfBibitem
\bibitem[Tang \latin{et~al.}(2017)Tang, Ma, S{\'a}nchez-D{\'\i}az, Ullbrich, Liu, Siegmund, Mischok, Leo, Campoy-Quiles, Li, \latin{et~al.} others]{tang2017polymer}
Tang,~Z.; Ma,~Z.; S{\'a}nchez-D{\'\i}az,~A.; Ullbrich,~S.; Liu,~Y.; Siegmund,~B.; Mischok,~A.; Leo,~K.; Campoy-Quiles,~M.; Li,~W. \latin{et~al.}  Polymer: fullerene bimolecular crystals for near-infrared spectroscopic photodetectors. \emph{Advanced Materials} \textbf{2017}, \emph{29}, 1702184\relax
\mciteBstWouldAddEndPuncttrue
\mciteSetBstMidEndSepPunct{\mcitedefaultmidpunct}
{\mcitedefaultendpunct}{\mcitedefaultseppunct}\relax
\EndOfBibitem
\bibitem[Wang \latin{et~al.}(2019)Wang, Ullbrich, Hou, Spoltore, Wang, Ma, Tang, and Vandewal]{wang2019organic}
Wang,~J.; Ullbrich,~S.; Hou,~J.-L.; Spoltore,~D.; Wang,~Q.; Ma,~Z.; Tang,~Z.; Vandewal,~K. Organic cavity photodetectors based on nanometer-thick active layers for tunable monochromatic spectral response. \emph{ACS Photonics} \textbf{2019}, \emph{6}, 1393--1399\relax
\mciteBstWouldAddEndPuncttrue
\mciteSetBstMidEndSepPunct{\mcitedefaultmidpunct}
{\mcitedefaultendpunct}{\mcitedefaultseppunct}\relax
\EndOfBibitem
\bibitem[Vanderspikken \latin{et~al.}(2022)Vanderspikken, Liu, Liu, Vandermeeren, Cardeynaels, Gielen, Van~Mele, Van~den Brande, Champagne, Vandewal, \latin{et~al.} others]{vanderspikken2022tuning}
Vanderspikken,~J.; Liu,~Q.; Liu,~Z.; Vandermeeren,~T.; Cardeynaels,~T.; Gielen,~S.; Van~Mele,~B.; Van~den Brande,~N.; Champagne,~B.; Vandewal,~K. \latin{et~al.}  Tuning electronic and morphological properties for high-performance wavelength-selective organic near-infrared cavity photodetectors. \emph{Advanced Functional Materials} \textbf{2022}, \emph{32}, 2108146\relax
\mciteBstWouldAddEndPuncttrue
\mciteSetBstMidEndSepPunct{\mcitedefaultmidpunct}
{\mcitedefaultendpunct}{\mcitedefaultseppunct}\relax
\EndOfBibitem
\bibitem[Xing \latin{et~al.}(2021)Xing, Nikolis, Kublitski, Guo, Jia, Wang, Spoltore, Vandewal, Kleemann, Benduhn, \latin{et~al.} others]{xing2021miniaturized}
Xing,~S.; Nikolis,~V.~C.; Kublitski,~J.; Guo,~E.; Jia,~X.; Wang,~Y.; Spoltore,~D.; Vandewal,~K.; Kleemann,~H.; Benduhn,~J. \latin{et~al.}  Miniaturized VIS-NIR spectrometers based on narrowband and tunable transmission cavity organic photodetectors with ultrahigh specific detectivity above $10^{14}$ jones. \emph{Advanced Materials} \textbf{2021}, \emph{33}, 2102967\relax
\mciteBstWouldAddEndPuncttrue
\mciteSetBstMidEndSepPunct{\mcitedefaultmidpunct}
{\mcitedefaultendpunct}{\mcitedefaultseppunct}\relax
\EndOfBibitem
\bibitem[Wang \latin{et~al.}(2021)Wang, Siegmund, Tang, Ma, Kublitski, Xing, Nikolis, Ullbrich, Li, Benduhn, \latin{et~al.} others]{wang2021stacked}
Wang,~Y.; Siegmund,~B.; Tang,~Z.; Ma,~Z.; Kublitski,~J.; Xing,~S.; Nikolis,~V.~C.; Ullbrich,~S.; Li,~Y.; Benduhn,~J. \latin{et~al.}  Stacked dual-wavelength near-infrared organic photodetectors. \emph{Advanced Optical Materials} \textbf{2021}, \emph{9}, 2001784\relax
\mciteBstWouldAddEndPuncttrue
\mciteSetBstMidEndSepPunct{\mcitedefaultmidpunct}
{\mcitedefaultendpunct}{\mcitedefaultseppunct}\relax
\EndOfBibitem
\bibitem[Yang \latin{et~al.}(2021)Yang, Huang, Li, Li, Sun, Lin, Wang, Ma, Vandewal, and Tang]{yang2021cavity}
Yang,~J.; Huang,~J.; Li,~R.; Li,~H.; Sun,~B.; Lin,~Q.; Wang,~M.; Ma,~Z.; Vandewal,~K.; Tang,~Z. Cavity-enhanced near-infrared organic photodetectors based on a conjugated polymer containing [1, 2, 5] selenadiazolo [3, 4-c] pyridine. \emph{Chemistry of Materials} \textbf{2021}, \emph{33}, 5147--5155\relax
\mciteBstWouldAddEndPuncttrue
\mciteSetBstMidEndSepPunct{\mcitedefaultmidpunct}
{\mcitedefaultendpunct}{\mcitedefaultseppunct}\relax
\EndOfBibitem
\bibitem[Palo and Daskalakis(2023)Palo, and Daskalakis]{Palo2023}
Palo,~E.; Daskalakis,~K.~S. {Prospects in Broadening the Application of Planar Solution‐Based Distributed Bragg Reflectors}. \emph{Adv. Mater. Interfaces} \textbf{2023}, 2202206\relax
\mciteBstWouldAddEndPuncttrue
\mciteSetBstMidEndSepPunct{\mcitedefaultmidpunct}
{\mcitedefaultendpunct}{\mcitedefaultseppunct}\relax
\EndOfBibitem
\bibitem[Frisk~Kockum \latin{et~al.}(2019)Frisk~Kockum, Miranowicz, De~Liberato, Savasta, and Nori]{frisk2019ultrastrong}
Frisk~Kockum,~A.; Miranowicz,~A.; De~Liberato,~S.; Savasta,~S.; Nori,~F. Ultrastrong coupling between light and matter. \emph{Nature Reviews Physics} \textbf{2019}, \emph{1}, 19--40\relax
\mciteBstWouldAddEndPuncttrue
\mciteSetBstMidEndSepPunct{\mcitedefaultmidpunct}
{\mcitedefaultendpunct}{\mcitedefaultseppunct}\relax
\EndOfBibitem
\bibitem[Garcia-Vidal \latin{et~al.}(2021)Garcia-Vidal, Ciuti, and Ebbesen]{garcia2021manipulating}
Garcia-Vidal,~F.~J.; Ciuti,~C.; Ebbesen,~T.~W. Manipulating matter by strong coupling to vacuum fields. \emph{Science} \textbf{2021}, \emph{373}, eabd0336\relax
\mciteBstWouldAddEndPuncttrue
\mciteSetBstMidEndSepPunct{\mcitedefaultmidpunct}
{\mcitedefaultendpunct}{\mcitedefaultseppunct}\relax
\EndOfBibitem
\bibitem[Coles \latin{et~al.}(2014)Coles, Somaschi, Michetti, Clark, Lagoudakis, Savvidis, and Lidzey]{coles2014polariton}
Coles,~D.~M.; Somaschi,~N.; Michetti,~P.; Clark,~C.; Lagoudakis,~P.~G.; Savvidis,~P.~G.; Lidzey,~D.~G. Polariton-mediated energy transfer between organic dyes in a strongly coupled optical microcavity. \emph{Nature materials} \textbf{2014}, \emph{13}, 712--719\relax
\mciteBstWouldAddEndPuncttrue
\mciteSetBstMidEndSepPunct{\mcitedefaultmidpunct}
{\mcitedefaultendpunct}{\mcitedefaultseppunct}\relax
\EndOfBibitem
\bibitem[Xiang \latin{et~al.}(2020)Xiang, Ribeiro, Du, Chen, Yang, Wang, Yuen-Zhou, and Xiong]{xiang2020intermolecular}
Xiang,~B.; Ribeiro,~R.~F.; Du,~M.; Chen,~L.; Yang,~Z.; Wang,~J.; Yuen-Zhou,~J.; Xiong,~W. Intermolecular vibrational energy transfer enabled by microcavity strong light--matter coupling. \emph{Science} \textbf{2020}, \emph{368}, 665--667\relax
\mciteBstWouldAddEndPuncttrue
\mciteSetBstMidEndSepPunct{\mcitedefaultmidpunct}
{\mcitedefaultendpunct}{\mcitedefaultseppunct}\relax
\EndOfBibitem
\bibitem[Wang \latin{et~al.}(2021)Wang, Hertzog, and B{\"o}rjesson]{wang2021polariton}
Wang,~M.; Hertzog,~M.; B{\"o}rjesson,~K. Polariton-assisted excitation energy channeling in organic heterojunctions. \emph{Nature communications} \textbf{2021}, \emph{12}, 1874\relax
\mciteBstWouldAddEndPuncttrue
\mciteSetBstMidEndSepPunct{\mcitedefaultmidpunct}
{\mcitedefaultendpunct}{\mcitedefaultseppunct}\relax
\EndOfBibitem
\bibitem[Balasubrahmaniyam \latin{et~al.}(2023)Balasubrahmaniyam, Simkhovich, Golombek, Sandik, Ankonina, and Schwartz]{balasubrahmaniyam2023enhanced}
Balasubrahmaniyam,~M.; Simkhovich,~A.; Golombek,~A.; Sandik,~G.; Ankonina,~G.; Schwartz,~T. From enhanced diffusion to ultrafast ballistic motion of hybrid light--matter excitations. \emph{Nature Materials} \textbf{2023}, \emph{22}, 338--344\relax
\mciteBstWouldAddEndPuncttrue
\mciteSetBstMidEndSepPunct{\mcitedefaultmidpunct}
{\mcitedefaultendpunct}{\mcitedefaultseppunct}\relax
\EndOfBibitem
\bibitem[Sokolovskii \latin{et~al.}(2023)Sokolovskii, Tichauer, Morozov, Feist, and Groenhof]{sokolovskii2023multi}
Sokolovskii,~I.; Tichauer,~R.~H.; Morozov,~D.; Feist,~J.; Groenhof,~G. Multi-scale molecular dynamics simulations of enhanced energy transfer in organic molecules under strong coupling. \emph{Nature Communications} \textbf{2023}, \emph{14}, 6613\relax
\mciteBstWouldAddEndPuncttrue
\mciteSetBstMidEndSepPunct{\mcitedefaultmidpunct}
{\mcitedefaultendpunct}{\mcitedefaultseppunct}\relax
\EndOfBibitem
\bibitem[Sandik \latin{et~al.}(2024)Sandik, Feist, Garc{\'\i}a-Vidal, and Schwartz]{sandik2024cavity}
Sandik,~G.; Feist,~J.; Garc{\'\i}a-Vidal,~F.~J.; Schwartz,~T. Cavity-enhanced energy transport in molecular systems. \emph{Nature Materials} \textbf{2024}, 1--12\relax
\mciteBstWouldAddEndPuncttrue
\mciteSetBstMidEndSepPunct{\mcitedefaultmidpunct}
{\mcitedefaultendpunct}{\mcitedefaultseppunct}\relax
\EndOfBibitem
\bibitem[Zhao \latin{et~al.}(2024)Zhao, Arneson, Fan, and Forrest]{zhao2024stable}
Zhao,~H.; Arneson,~C.~E.; Fan,~D.; Forrest,~S.~R. Stable blue phosphorescent organic LEDs that use polariton-enhanced Purcell effects. \emph{Nature} \textbf{2024}, \emph{626}, 300--305\relax
\mciteBstWouldAddEndPuncttrue
\mciteSetBstMidEndSepPunct{\mcitedefaultmidpunct}
{\mcitedefaultendpunct}{\mcitedefaultseppunct}\relax
\EndOfBibitem
\bibitem[Arneson \latin{et~al.}(2024)Arneson, Zhao, and Forrest]{arneson2024color}
Arneson,~C.~E.; Zhao,~H.; Forrest,~S.~R. Color-Stable, All-Phosphorescent White Organic Light Emitting Diodes Using the Polariton-Enhanced Purcell Effect. \emph{Advanced Functional Materials} \textbf{2024}, 2410741\relax
\mciteBstWouldAddEndPuncttrue
\mciteSetBstMidEndSepPunct{\mcitedefaultmidpunct}
{\mcitedefaultendpunct}{\mcitedefaultseppunct}\relax
\EndOfBibitem
\bibitem[Qureshi \latin{et~al.}(2024)Qureshi, Papachatzakis, Abdelmagid, Salom{\"a}ki, M{\"a}kil{\"a}, Siltanen, and Daskalakis]{qureshi2024giant}
Qureshi,~H.~A.; Papachatzakis,~M.~A.; Abdelmagid,~A.~G.; Salom{\"a}ki,~M.; M{\"a}kil{\"a},~E.; Siltanen,~O.; Daskalakis,~K.~S. Giant Rabi splitting and polariton photoluminescence in an all solution-deposited dielectric microcavity. \emph{arXiv preprint arXiv:2410.19392} \textbf{2024}, \relax
\mciteBstWouldAddEndPunctfalse
\mciteSetBstMidEndSepPunct{\mcitedefaultmidpunct}
{}{\mcitedefaultseppunct}\relax
\EndOfBibitem
\bibitem[Deng \latin{et~al.}(2010)Deng, Haug, and Yamamoto]{deng2010exciton}
Deng,~H.; Haug,~H.; Yamamoto,~Y. Exciton-polariton bose-einstein condensation. \emph{Reviews of modern physics} \textbf{2010}, \emph{82}, 1489--1537\relax
\mciteBstWouldAddEndPuncttrue
\mciteSetBstMidEndSepPunct{\mcitedefaultmidpunct}
{\mcitedefaultendpunct}{\mcitedefaultseppunct}\relax
\EndOfBibitem
\bibitem[Eizner \latin{et~al.}(2018)Eizner, Brodeur, Barachati, Sridharan, and K{\'e}na-Cohen]{eizner2018organic}
Eizner,~E.; Brodeur,~J.; Barachati,~F.; Sridharan,~A.; K{\'e}na-Cohen,~S. Organic photodiodes with an extended responsivity using ultrastrong light--matter coupling. \emph{ACS Photonics} \textbf{2018}, \emph{5}, 2921--2927\relax
\mciteBstWouldAddEndPuncttrue
\mciteSetBstMidEndSepPunct{\mcitedefaultmidpunct}
{\mcitedefaultendpunct}{\mcitedefaultseppunct}\relax
\EndOfBibitem
\bibitem[Mischok \latin{et~al.}(2020)Mischok, L{\"u}ttgens, Berger, Hillebrandt, Tenopala-Carmona, Kwon, Murawski, Siegmund, Zaumseil, and Gather]{mischok2020spectroscopic}
Mischok,~A.; L{\"u}ttgens,~J.; Berger,~F.; Hillebrandt,~S.; Tenopala-Carmona,~F.; Kwon,~S.; Murawski,~C.; Siegmund,~B.; Zaumseil,~J.; Gather,~M.~C. Spectroscopic near-infrared photodetectors enabled by strong light--matter coupling in (6, 5) single-walled carbon nanotubes. \emph{The Journal of Chemical Physics} \textbf{2020}, \emph{153}\relax
\mciteBstWouldAddEndPuncttrue
\mciteSetBstMidEndSepPunct{\mcitedefaultmidpunct}
{\mcitedefaultendpunct}{\mcitedefaultseppunct}\relax
\EndOfBibitem
\bibitem[Mischok \latin{et~al.}(2024)Mischok, Siegmund, Le~Roux, Hillebrandt, Vandewal, and Gather]{mischok2024breaking}
Mischok,~A.; Siegmund,~B.; Le~Roux,~F.; Hillebrandt,~S.; Vandewal,~K.; Gather,~M.~C. Breaking the angular dispersion limit in thin film optics by ultra-strong light-matter coupling. \emph{Nature Communications} \textbf{2024}, \emph{15}, 1--10\relax
\mciteBstWouldAddEndPuncttrue
\mciteSetBstMidEndSepPunct{\mcitedefaultmidpunct}
{\mcitedefaultendpunct}{\mcitedefaultseppunct}\relax
\EndOfBibitem
\bibitem[Mischok \latin{et~al.}(2023)Mischok, Hillebrandt, Kwon, and Gather]{mischok2023highly}
Mischok,~A.; Hillebrandt,~S.; Kwon,~S.; Gather,~M.~C. Highly efficient polaritonic light-emitting diodes with angle-independent narrowband emission. \emph{Nature Photonics} \textbf{2023}, \emph{17}, 393--400\relax
\mciteBstWouldAddEndPuncttrue
\mciteSetBstMidEndSepPunct{\mcitedefaultmidpunct}
{\mcitedefaultendpunct}{\mcitedefaultseppunct}\relax
\EndOfBibitem
\bibitem[De \latin{et~al.}(2024)De, Zhao, Yin, Gu, Long, Huang, Cao, An, Liao, Fu, \latin{et~al.} others]{de2024organic}
De,~J.; Zhao,~R.; Yin,~F.; Gu,~C.; Long,~T.; Huang,~H.; Cao,~X.; An,~C.; Liao,~B.; Fu,~H. \latin{et~al.}  Organic polaritonic light-emitting diodes with high luminance and color purity toward laser displays. \emph{Light: Science \& Applications} \textbf{2024}, \emph{13}, 191\relax
\mciteBstWouldAddEndPuncttrue
\mciteSetBstMidEndSepPunct{\mcitedefaultmidpunct}
{\mcitedefaultendpunct}{\mcitedefaultseppunct}\relax
\EndOfBibitem
\bibitem[Souza \latin{et~al.}(2022)Souza, Benatto, Candiotto, Roman, and Koehler]{souza2022binding}
Souza,~J.; Benatto,~L.; Candiotto,~G.; Roman,~L.; Koehler,~M. Binding energy of triplet excitons in nonfullerene acceptors: the effects of fluorination and chlorination. \emph{The Journal of Physical Chemistry A} \textbf{2022}, \emph{126}, 1393--1402\relax
\mciteBstWouldAddEndPuncttrue
\mciteSetBstMidEndSepPunct{\mcitedefaultmidpunct}
{\mcitedefaultendpunct}{\mcitedefaultseppunct}\relax
\EndOfBibitem
\bibitem[Song \latin{et~al.}(2018)Song, Gasparini, Ye, Yao, Hou, Ade, and Baran]{song2018controlling}
Song,~X.; Gasparini,~N.; Ye,~L.; Yao,~H.; Hou,~J.; Ade,~H.; Baran,~D. Controlling blend morphology for ultrahigh current density in nonfullerene acceptor-based organic solar cells. \emph{ACS Energy Letters} \textbf{2018}, \emph{3}, 669--676\relax
\mciteBstWouldAddEndPuncttrue
\mciteSetBstMidEndSepPunct{\mcitedefaultmidpunct}
{\mcitedefaultendpunct}{\mcitedefaultseppunct}\relax
\EndOfBibitem
\bibitem[Bhuyan \latin{et~al.}(2023)Bhuyan, Mony, Kotov, Castellanos, G{\'o}mez~Rivas, Shegai, and B{\"o}rjesson]{bhuyan2023rise}
Bhuyan,~R.; Mony,~J.; Kotov,~O.; Castellanos,~G.~W.; G{\'o}mez~Rivas,~J.; Shegai,~T.~O.; B{\"o}rjesson,~K. The rise and current status of polaritonic photochemistry and photophysics. \emph{Chemical Reviews} \textbf{2023}, \emph{123}, 10877--10919\relax
\mciteBstWouldAddEndPuncttrue
\mciteSetBstMidEndSepPunct{\mcitedefaultmidpunct}
{\mcitedefaultendpunct}{\mcitedefaultseppunct}\relax
\EndOfBibitem
\bibitem[Saleem \latin{et~al.}(2021)Saleem, Farhat, Khera, Langer, and Iqbal]{saleem2021designing}
Saleem,~R.; Farhat,~A.; Khera,~R.~A.; Langer,~P.; Iqbal,~J. Designing of small molecule non-fullerene acceptors with cyanobenzene core for photovoltaic application. \emph{Computational and Theoretical Chemistry} \textbf{2021}, \emph{1197}, 113154\relax
\mciteBstWouldAddEndPuncttrue
\mciteSetBstMidEndSepPunct{\mcitedefaultmidpunct}
{\mcitedefaultendpunct}{\mcitedefaultseppunct}\relax
\EndOfBibitem
\bibitem[Zhong \latin{et~al.}(2024)Zhong, Liu, and You]{zhong2024general}
Zhong,~X.; Liu,~S.; You,~W. A general and mild synthetic method for fused-ring electronic acceptors. \emph{Science Advances} \textbf{2024}, \emph{10}, eadp8150\relax
\mciteBstWouldAddEndPuncttrue
\mciteSetBstMidEndSepPunct{\mcitedefaultmidpunct}
{\mcitedefaultendpunct}{\mcitedefaultseppunct}\relax
\EndOfBibitem
\bibitem[Wadsworth \latin{et~al.}(2019)Wadsworth, Moser, Marks, Little, Gasparini, Brabec, Baran, and McCulloch]{wadsworth2019critical}
Wadsworth,~A.; Moser,~M.; Marks,~A.; Little,~M.~S.; Gasparini,~N.; Brabec,~C.~J.; Baran,~D.; McCulloch,~I. Critical review of the molecular design progress in non-fullerene electron acceptors towards commercially viable organic solar cells. \emph{Chemical Society Reviews} \textbf{2019}, \emph{48}, 1596--1625\relax
\mciteBstWouldAddEndPuncttrue
\mciteSetBstMidEndSepPunct{\mcitedefaultmidpunct}
{\mcitedefaultendpunct}{\mcitedefaultseppunct}\relax
\EndOfBibitem
\bibitem[Moustafa \latin{et~al.}(2023)Moustafa, M{\'e}ndez, S{\'a}nchez, Pallar{\`e}s, Palomares, and Marsal]{moustafa2023thermal}
Moustafa,~E.; M{\'e}ndez,~M.; S{\'a}nchez,~J.~G.; Pallar{\`e}s,~J.; Palomares,~E.; Marsal,~L.~F. Thermal Activation of PEDOT: PSS/PM6: Y7 Based Films Leads to Unprecedent High Short-Circuit Current Density in Nonfullerene Organic Photovoltaics. \emph{Advanced Energy Materials} \textbf{2023}, \emph{13}, 2203241\relax
\mciteBstWouldAddEndPuncttrue
\mciteSetBstMidEndSepPunct{\mcitedefaultmidpunct}
{\mcitedefaultendpunct}{\mcitedefaultseppunct}\relax
\EndOfBibitem
\bibitem[Zhu \latin{et~al.}(2022)Zhu, Zhang, Xu, Li, Yan, Zhou, Zhong, Hao, Song, Xue, \latin{et~al.} others]{zhu2022single}
Zhu,~L.; Zhang,~M.; Xu,~J.; Li,~C.; Yan,~J.; Zhou,~G.; Zhong,~W.; Hao,~T.; Song,~J.; Xue,~X. \latin{et~al.}  Single-junction organic solar cells with over 19\% efficiency enabled by a refined double-fibril network morphology. \emph{Nature Materials} \textbf{2022}, \emph{21}, 656--663\relax
\mciteBstWouldAddEndPuncttrue
\mciteSetBstMidEndSepPunct{\mcitedefaultmidpunct}
{\mcitedefaultendpunct}{\mcitedefaultseppunct}\relax
\EndOfBibitem
\bibitem[Chen \latin{et~al.}(2023)Chen, Jeong, Tian, Zhang, Naphade, Alsufyani, Zhang, Griggs, Hu, Barlow, \latin{et~al.} others]{chen202319}
Chen,~H.; Jeong,~S.~Y.; Tian,~J.; Zhang,~Y.; Naphade,~D.~R.; Alsufyani,~M.; Zhang,~W.; Griggs,~S.; Hu,~H.; Barlow,~S. \latin{et~al.}  A 19\% efficient and stable organic photovoltaic device enabled by a guest nonfullerene acceptor with fibril-like morphology. \emph{Energy \& Environmental Science} \textbf{2023}, \emph{16}, 1062--1070\relax
\mciteBstWouldAddEndPuncttrue
\mciteSetBstMidEndSepPunct{\mcitedefaultmidpunct}
{\mcitedefaultendpunct}{\mcitedefaultseppunct}\relax
\EndOfBibitem
\bibitem[Sun \latin{et~al.}(2024)Sun, Wang, Guo, Xiao, Liu, Chen, Xia, Gan, Cheng, Zhou, \latin{et~al.} others]{sun2024pi}
Sun,~Y.; Wang,~L.; Guo,~C.; Xiao,~J.; Liu,~C.; Chen,~C.; Xia,~W.; Gan,~Z.; Cheng,~J.; Zhou,~J. \latin{et~al.}  $\pi$-Extended Nonfullerene Acceptor for Compressed Molecular Packing in Organic Solar Cells To Achieve over 20\% Efficiency. \emph{Journal of the American Chemical Society} \textbf{2024}, \emph{146}, 12011--12019\relax
\mciteBstWouldAddEndPuncttrue
\mciteSetBstMidEndSepPunct{\mcitedefaultmidpunct}
{\mcitedefaultendpunct}{\mcitedefaultseppunct}\relax
\EndOfBibitem
\bibitem[Yang \latin{et~al.}(2021)Yang, Qiu, Georgitzikis, Simoen, Serron, Lee, Lieberman, Cheyns, Malinowski, Genoe, \latin{et~al.} others]{yang2021mitigating}
Yang,~W.; Qiu,~W.; Georgitzikis,~E.; Simoen,~E.; Serron,~J.; Lee,~J.; Lieberman,~I.; Cheyns,~D.; Malinowski,~P.; Genoe,~J. \latin{et~al.}  Mitigating dark current for high-performance near-infrared organic photodiodes via charge blocking and defect passivation. \emph{ACS Applied Materials \& Interfaces} \textbf{2021}, \emph{13}, 16766--16774\relax
\mciteBstWouldAddEndPuncttrue
\mciteSetBstMidEndSepPunct{\mcitedefaultmidpunct}
{\mcitedefaultendpunct}{\mcitedefaultseppunct}\relax
\EndOfBibitem
\bibitem[Siddik \latin{et~al.}(2023)Siddik, Georgitzikis, Hermans, Kang, Kim, Pejovic, Lieberman, Malinowski, Kadashchuk, Genoe, \latin{et~al.} others]{siddik2023interface}
Siddik,~A.~B.; Georgitzikis,~E.; Hermans,~Y.; Kang,~J.; Kim,~J.~H.; Pejovic,~V.; Lieberman,~I.; Malinowski,~P.~E.; Kadashchuk,~A.; Genoe,~J. \latin{et~al.}  Interface-engineered organic near-infrared photodetector for imaging applications. \emph{ACS Applied Materials \& Interfaces} \textbf{2023}, \emph{15}, 30534--30542\relax
\mciteBstWouldAddEndPuncttrue
\mciteSetBstMidEndSepPunct{\mcitedefaultmidpunct}
{\mcitedefaultendpunct}{\mcitedefaultseppunct}\relax
\EndOfBibitem
\bibitem[Park \latin{et~al.}(2023)Park, Labanti, Pacalaj, Lee, Dong, Chin, Luke, Ryu, Minami, Yun, \latin{et~al.} others]{park2023state}
Park,~S.~Y.; Labanti,~C.; Pacalaj,~R.~A.; Lee,~T.~H.; Dong,~Y.; Chin,~Y.-C.; Luke,~J.; Ryu,~G.; Minami,~D.; Yun,~S. \latin{et~al.}  The State-of-the-Art Solution-Processed Single Component Organic Photodetectors Achieved by Strong Quenching of Intermolecular Emissive State and High Quadrupole Moment in Non-Fullerene Acceptors. \emph{Advanced Materials} \textbf{2023}, \emph{35}, 2306655\relax
\mciteBstWouldAddEndPuncttrue
\mciteSetBstMidEndSepPunct{\mcitedefaultmidpunct}
{\mcitedefaultendpunct}{\mcitedefaultseppunct}\relax
\EndOfBibitem
\bibitem[Luong \latin{et~al.}(2024)Luong, Kaiyasuan, Yi, Chae, Kim, Panoy, Kim, Promarak, Miyata, Nakayama, \latin{et~al.} others]{luong2024highly}
Luong,~H.~M.; Kaiyasuan,~C.; Yi,~A.; Chae,~S.; Kim,~B.~M.; Panoy,~P.; Kim,~H.~J.; Promarak,~V.; Miyata,~Y.; Nakayama,~H. \latin{et~al.}  Highly Sensitive Resonance-Enhanced Organic Photodetectors for Shortwave Infrared Sensing. \emph{ACS Energy Letters} \textbf{2024}, \emph{9}, 1446--1454\relax
\mciteBstWouldAddEndPuncttrue
\mciteSetBstMidEndSepPunct{\mcitedefaultmidpunct}
{\mcitedefaultendpunct}{\mcitedefaultseppunct}\relax
\EndOfBibitem
\bibitem[Abdelmagid \latin{et~al.}(2024)Abdelmagid, Qureshi, Papachatzakis, Siltanen, Kumar, Ashokan, Salman, Luoma, and Daskalakis]{abdelmagid2024identifying}
Abdelmagid,~A.~G.; Qureshi,~H.~A.; Papachatzakis,~M.~A.; Siltanen,~O.; Kumar,~M.; Ashokan,~A.; Salman,~S.; Luoma,~K.; Daskalakis,~K.~S. Identifying the origin of delayed electroluminescence in a polariton organic light-emitting diode. \emph{Nanophotonics} \textbf{2024}, \emph{13}, 2565--2573\relax
\mciteBstWouldAddEndPuncttrue
\mciteSetBstMidEndSepPunct{\mcitedefaultmidpunct}
{\mcitedefaultendpunct}{\mcitedefaultseppunct}\relax
\EndOfBibitem
\bibitem[Hu \latin{et~al.}(2024)Hu, Qiao, Nodari, He, Asatryan, Rimmele, Chen, Mart{\'\i}n, Gasparini, and Heeney]{hu2024remarkable}
Hu,~X.; Qiao,~Z.; Nodari,~D.; He,~Q.; Asatryan,~J.; Rimmele,~M.; Chen,~Z.; Mart{\'\i}n,~J.; Gasparini,~N.; Heeney,~M. Remarkable Isomer Effect on the Performance of Fully Non-Fused Non-Fullerene Acceptors in Near-Infrared Organic Photodetectors. \emph{Advanced Optical Materials} \textbf{2024}, \emph{12}, 2302210\relax
\mciteBstWouldAddEndPuncttrue
\mciteSetBstMidEndSepPunct{\mcitedefaultmidpunct}
{\mcitedefaultendpunct}{\mcitedefaultseppunct}\relax
\EndOfBibitem
\bibitem[Du \latin{et~al.}(2024)Du, Luong, Sabury, Jones, Zhu, Panoy, Chae, Yi, Kim, Xiao, \latin{et~al.} others]{du2024high}
Du,~Z.; Luong,~H.~M.; Sabury,~S.; Jones,~A.~L.; Zhu,~Z.; Panoy,~P.; Chae,~S.; Yi,~A.; Kim,~H.~J.; Xiao,~S. \latin{et~al.}  High-Performance Wearable Organic Photodetectors by Molecular Design and Green Solvent Processing for Pulse Oximetry and Photoplethysmography. \emph{Advanced Materials} \textbf{2024}, \emph{36}, 2310478\relax
\mciteBstWouldAddEndPuncttrue
\mciteSetBstMidEndSepPunct{\mcitedefaultmidpunct}
{\mcitedefaultendpunct}{\mcitedefaultseppunct}\relax
\EndOfBibitem
\bibitem[Jacoutot \latin{et~al.}(2023)Jacoutot, Scaccabarozzi, Nodari, Panidi, Qiao, Schiza, Nega, Dimitrakopoulou-Strauss, Gregoriou, Heeney, \latin{et~al.} others]{jacoutot2023enhanced}
Jacoutot,~P.; Scaccabarozzi,~A.~D.; Nodari,~D.; Panidi,~J.; Qiao,~Z.; Schiza,~A.; Nega,~A.~D.; Dimitrakopoulou-Strauss,~A.; Gregoriou,~V.~G.; Heeney,~M. \latin{et~al.}  Enhanced sub-1 eV detection in organic photodetectors through tuning polymer energetics and microstructure. \emph{Science Advances} \textbf{2023}, \emph{9}, eadh2694\relax
\mciteBstWouldAddEndPuncttrue
\mciteSetBstMidEndSepPunct{\mcitedefaultmidpunct}
{\mcitedefaultendpunct}{\mcitedefaultseppunct}\relax
\EndOfBibitem
\end{mcitethebibliography}

\newpage

\newpage
\setcounter{equation}{0}
\setcounter{figure}{0}
\setcounter{table}{0}
\setcounter{page}{1}
\makeatletter
\renewcommand{\theequation}{S\arabic{equation}}
\renewcommand{\thefigure}{S\arabic{figure}}

\begin{center}
\textbf{\Large Supplementary Information}\\
\end{center}
\begin{center}
\textbf{\Large{Polaritons in non-fullerene acceptors for high responsivity angle-independent organic narrowband infrared photodiodes}}\\
\end{center}
\noindent
Ahmed Gaber Abdelmagid, Zhuoran Qiao, Boudewijn Coenegracht, Gaon
Yu, Hassan A. Qureshi, Thomas D. Anthopoulos, Nicola Gasparini, and Konstantinos S. Daskalakis\\
\noindent
Corresponding authors: 
ahmed.abdelmagid@utu.fi; konstantinos.daskalakis@utu.fi
\section*{Contents}

\begin{figure}[h!]
\vspace{0pt}
\centering
\includegraphics[width= 0.75\textwidth]{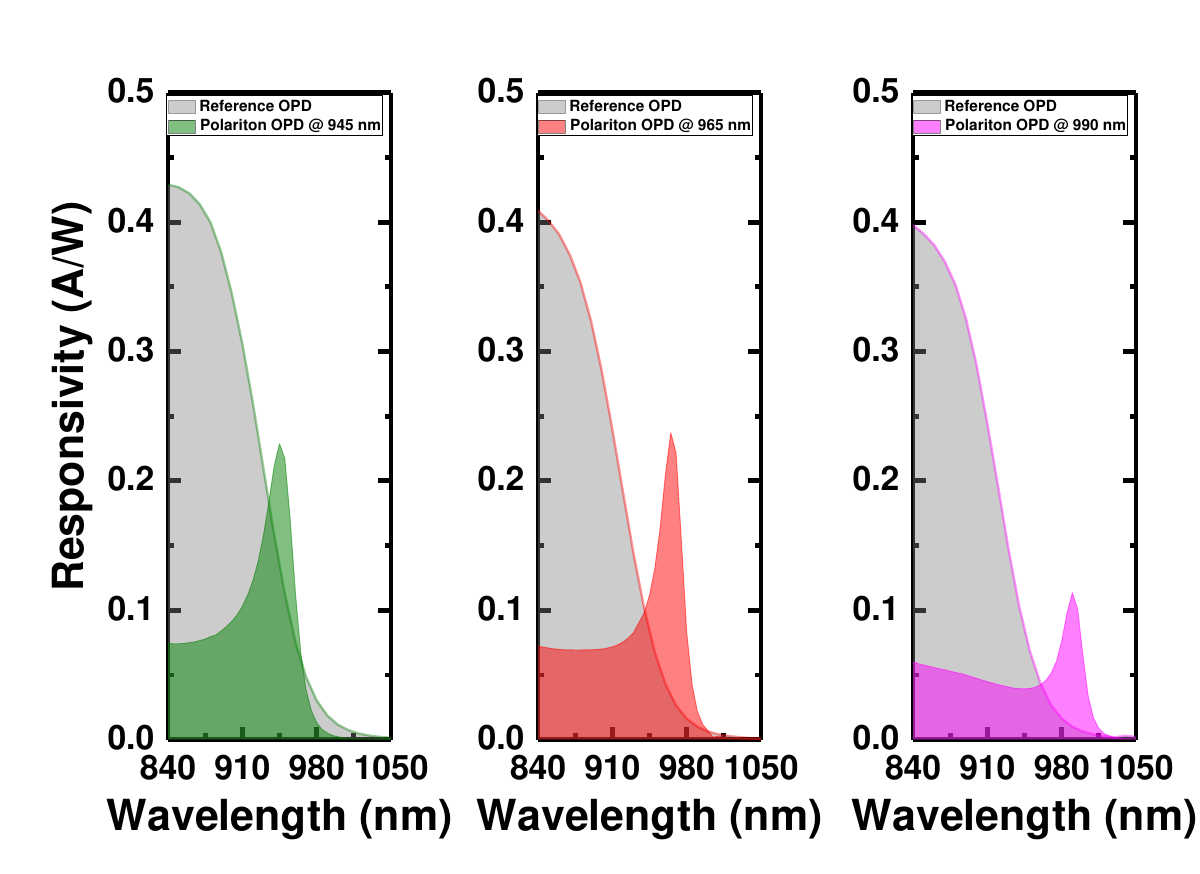}
\vspace{-12pt}
\caption{The responsivity spectra of polariton OPDs with device resonance at 945~nm (olive), 965~nm (red), 990~nm (magenta), compared to a reference OPD (grey) at -2~V.}
\label{Ref vs Polariton OPDs}
\vspace{0pt}
\end{figure}

\begin{figure}[h!]
\vspace{0pt}
\centering
\includegraphics[width=0.7\textwidth]{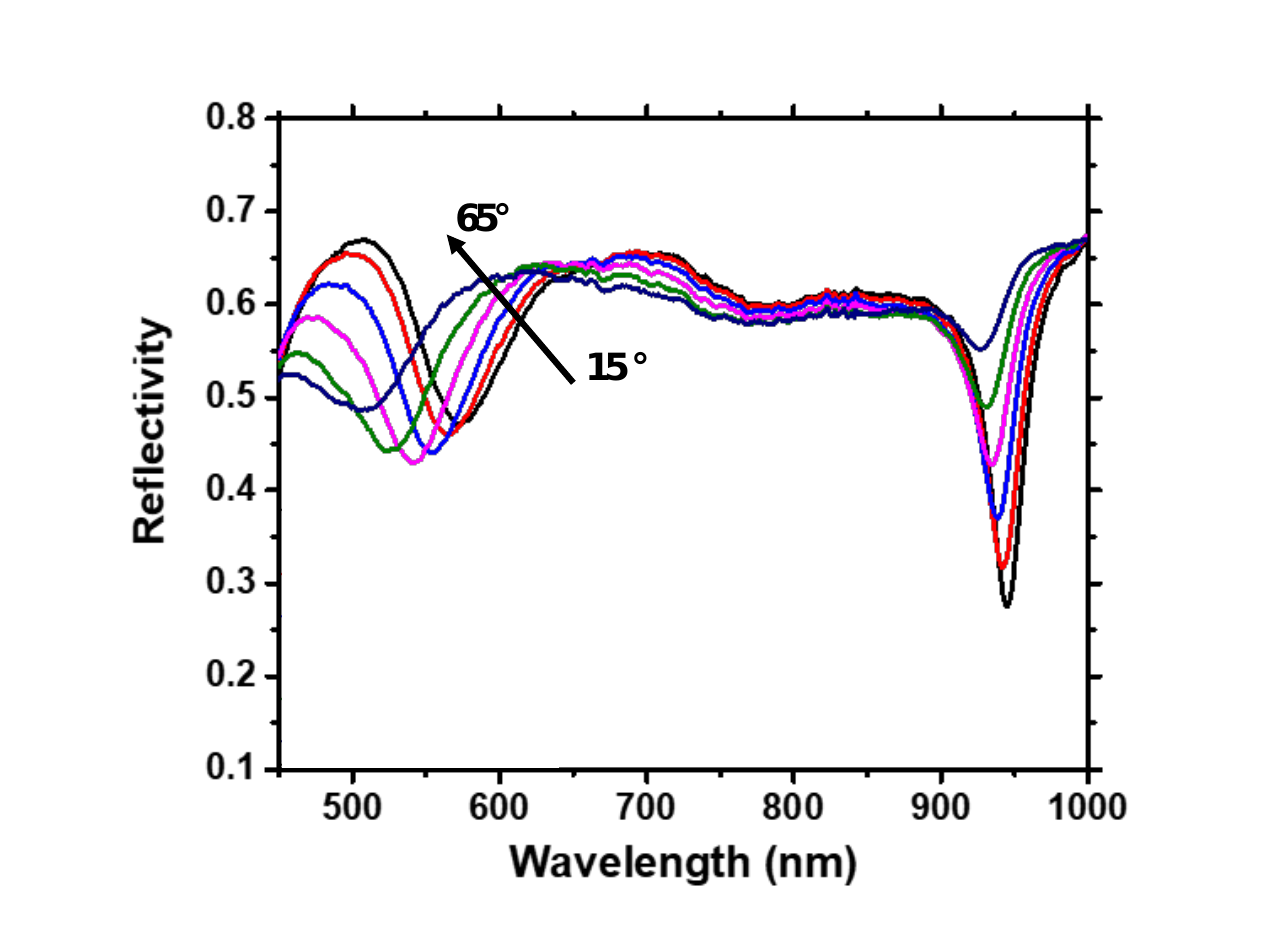}
\vspace{-12pt}
\caption{Polariton characteristics. Angle-resolved reflectivity spectra of a polariton OPD with device resonance at 945~nm. Angles were taken from 15° to 65° with an interval of 10°.}
\label{POPD ref spectra}
\vspace{0pt}
\end{figure}

\begin{figure}[h!]
\vspace{0pt}
\centering
\includegraphics[width=0.7\textwidth]{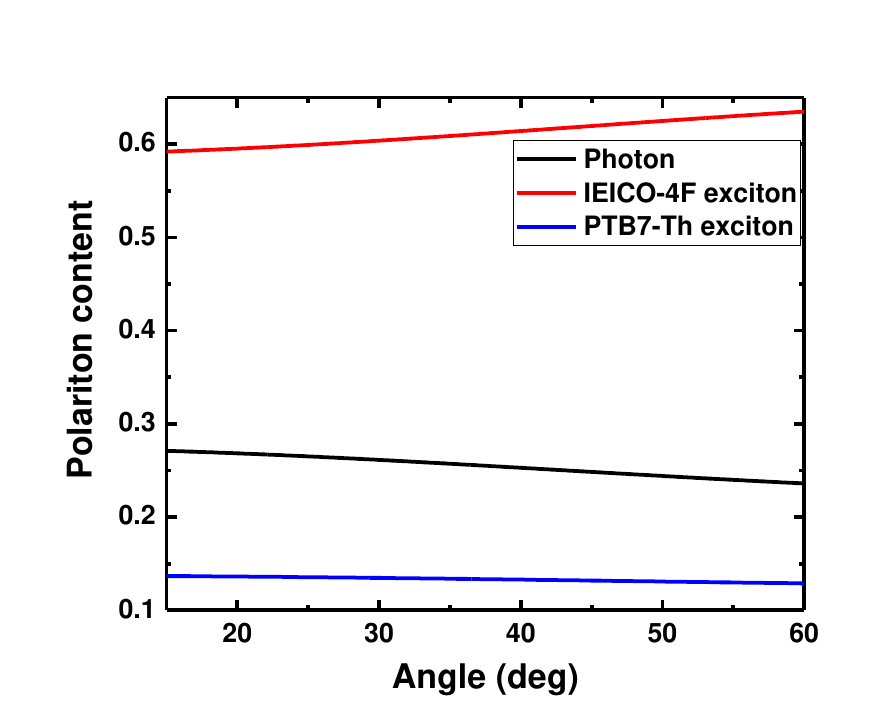}
\vspace{-12pt}
\caption{Polariton characteristics. Excitons and photon contents of the LP branch for polariton OPD with device resonance at 945~nm.}
\label{POPD945 fraction}
\vspace{0pt}
\end{figure}

\begin{figure}[h!]
\vspace{0pt}
\centering
\includegraphics[width=\linewidth]{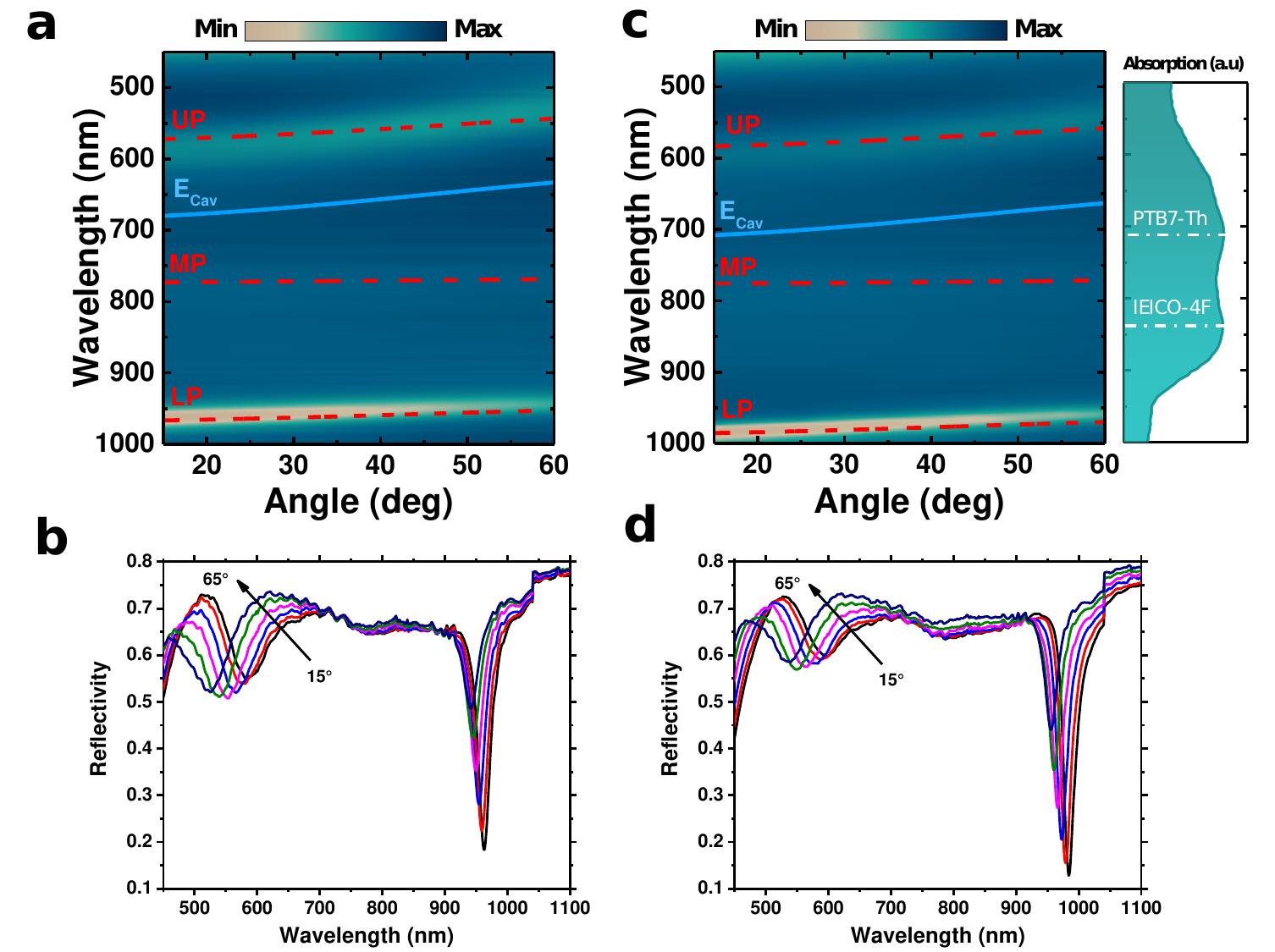}
\vspace{-12pt}
\caption{Polariton characteristics. Angle-resolved reflectivity a) map and 
 b) spectra of polariton OPD with device resonance of 965~nm. Angle-resolved reflectivity c) map and d) spectra of polariton OPD with device resonance of 990~nm. The solid blue line is the cavity energy dispersion, and the dashed red lines are fitted polariton dispersions. Besides is the absorption of the active blend and the dashed white line is the molecular exciton energies of PTB7-Th and IEICO-4F.}
\label{Device ref analysis}
\vspace{0pt}
\end{figure}

\begin{figure}[h!]
\vspace{0pt}
\centering
\includegraphics[width=\linewidth]{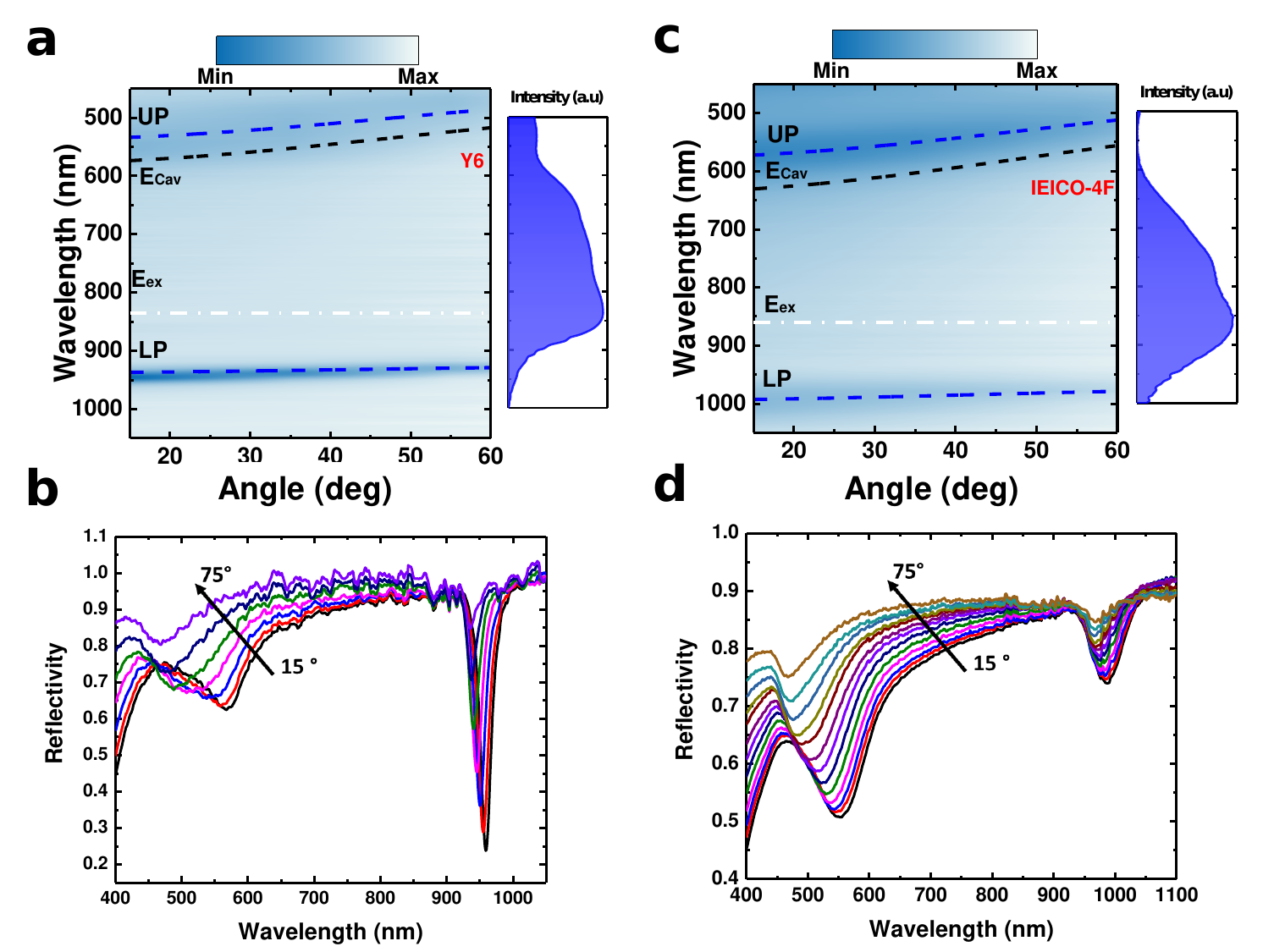}
\vspace{-12pt}
\caption{Polariton characteristics. Angle-resolved reflectivity a) map and b) spectra of the thin film microcavities of Y6. Angle-resolved reflectivity c) map and d) spectra of the thin film microcavities of IEICO-4F. The dashed white line is the molecular exciton energies, the dashed black line is the cavity energy dispersion, and the dashed blue lines are fitted polariton dispersions. Besides is the molecular absorption.}
\label{MC NFAs}
\vspace{0pt}
\end{figure}

\begin{figure}[h!]
\vspace{0pt}
\centering
\includegraphics[width=0.7\textwidth]{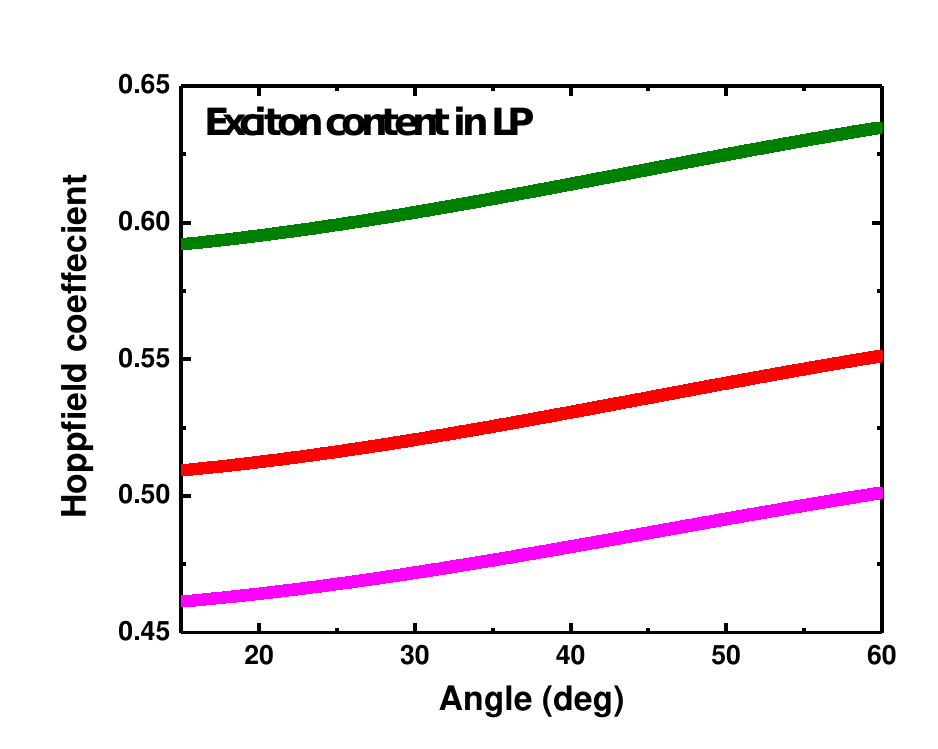}
\vspace{-12pt}
\caption{IEICO-4F exciton content in the LP for polariton OPD with device resonance at 945~nm (olive), 965~nm (red), 990~nm (magenta).}
\label{LP ex fraction}
\vspace{0pt}
\end{figure}

\begin{figure}[h!]
\vspace{0pt}
\centering
\includegraphics[width= 0.5\textwidth]{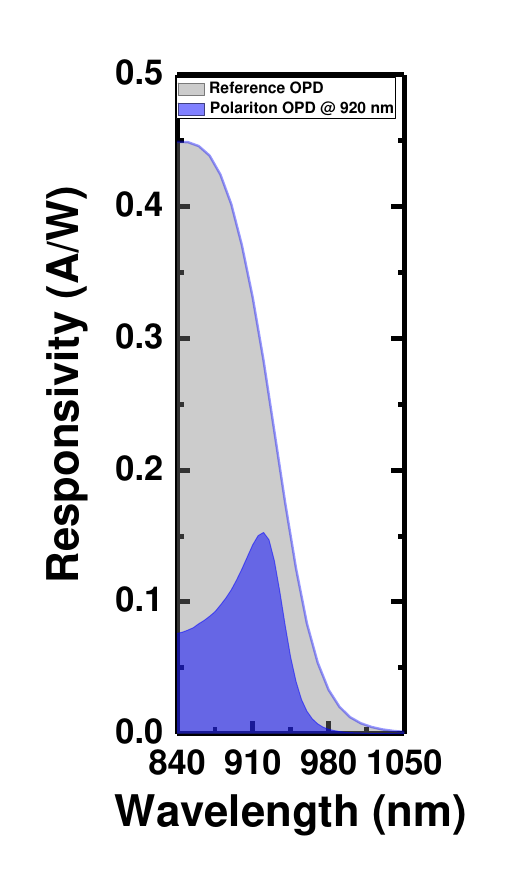}
\vspace{-12pt}
\caption{The responsivity spectra of a polariton OPDs with device resonance at 920~nm (blue) compared to a reference OPD (grey) at -2~V.}
\label{Polariton OPD920}
\vspace{0pt}
\end{figure}

\begin{figure}[h!]
\vspace{0pt}
\centering
\includegraphics[width=0.7\textwidth]{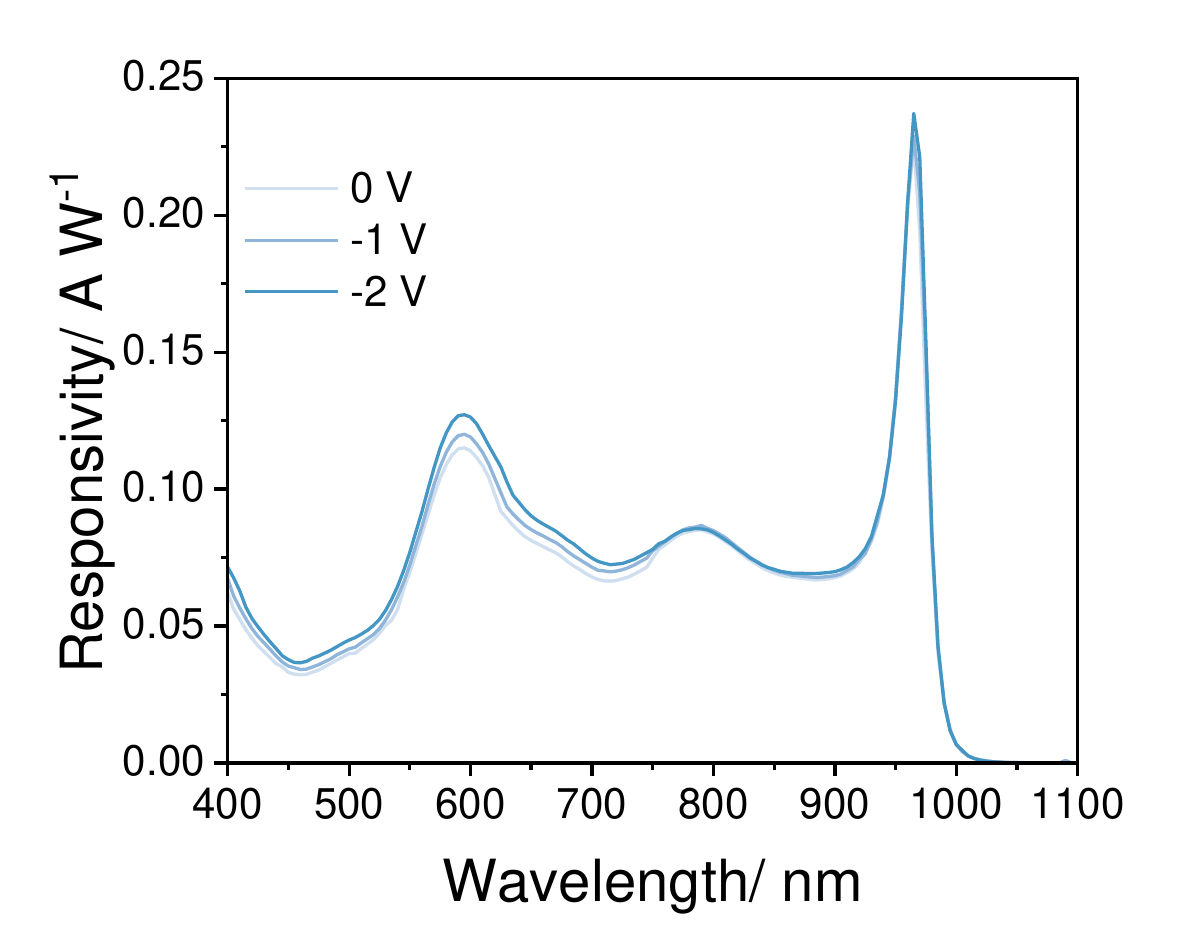}
\vspace{-12pt}
\caption{Resposnivity spectra of Polariton OPD with device resonance at 965~nm at different bias voltages.}
\label{POPD Rs spectra}
\vspace{0pt}
\end{figure}

\begin{figure}[h!]
\vspace{0pt}
\centering
\includegraphics[width=0.7\textwidth]{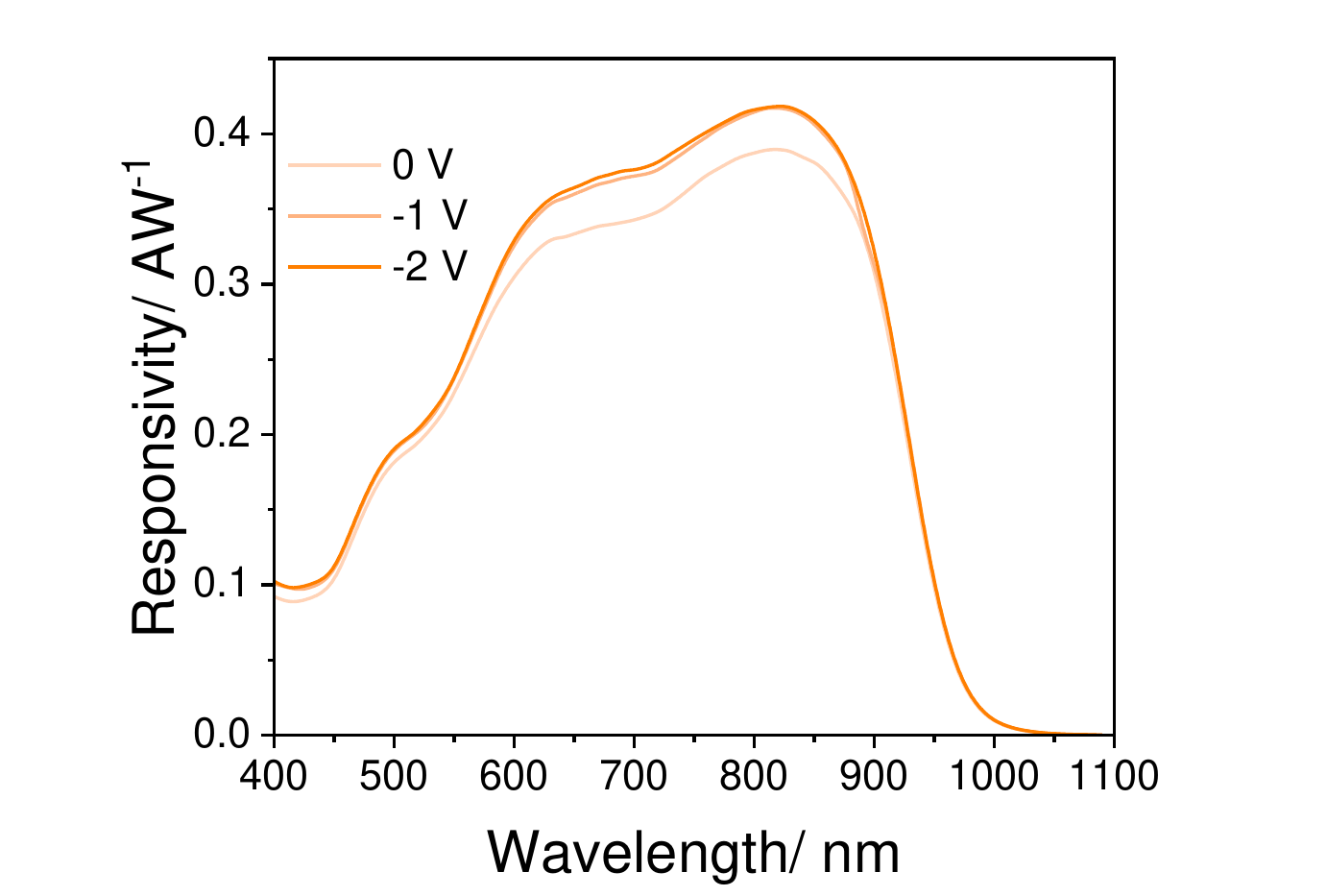}
\vspace{-12pt}
\caption{Responsivity spectra of the reference device at different bias voltages.}
\label{ref Rs spectra}
\vspace{0pt}
\end{figure}

\begin{figure}[h!]
\vspace{0pt}
\centering
\includegraphics[width=\linewidth]{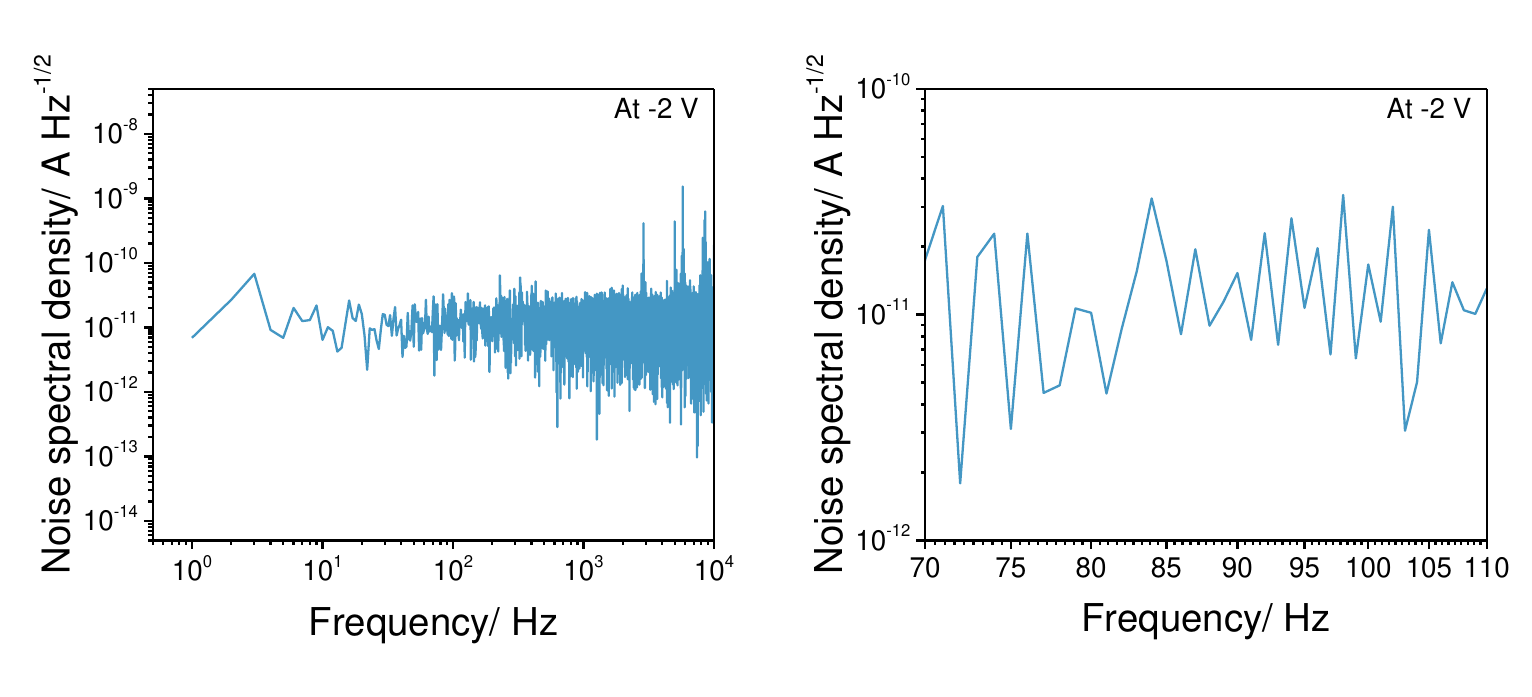}
\vspace{-12pt}
\caption{Noise spectral density of polariton OPD with device resonance at 965~nm at -2~V. }
\label{Polariton OPD noise}
\vspace{0pt}
\end{figure}

\begin{figure}[h!]
\vspace{0pt}
\centering
\includegraphics[width=\linewidth]{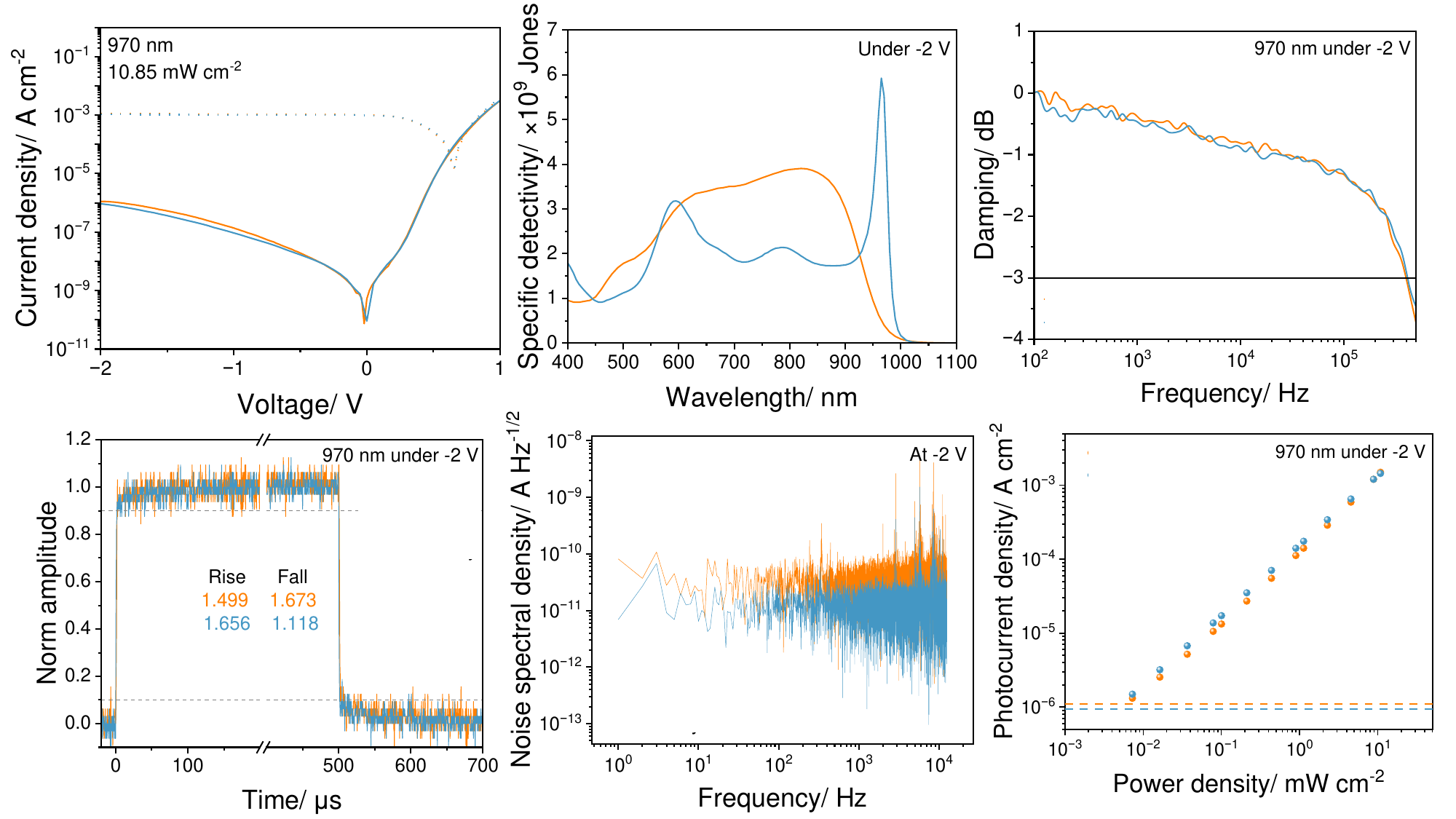}
\vspace{-12pt}
\caption{Comparison of key metrics of the polariton OPD (blue) and a reference device (orange).}
\label{comparison}
\vspace{0pt}
\end{figure}

\textbf{Supplementary Figure \ref{Ref vs Polariton OPDs}}. The responsivity spectra of the polariton OPDs compared to a reference OPD.\\
\textbf{Supplementary Figure \ref{POPD ref spectra}}. Angle-resolved reflectivity spectra of a polariton OPD (945~nm).\\
\textbf{Supplementary Figure \ref{POPD945 fraction}}. Excitons and photon contents.\\
\textbf{Supplementary Figure \ref{Device ref analysis}}. Angle-resolved reflectivity of polariton OPDs.\\
\textbf{Supplementary Figure \ref{MC NFAs}}. Angle-resolved reflectivity of NFAs microcavities.\\
\textbf{Supplementary Figure \ref{LP ex fraction}}. IEICO-4F exciton content in the LP for the polariton OPDs.\\
\textbf{Supplementary Figure \ref{POPD Rs spectra}}. Responsivity spectra of a Polariton OPD (965~nm) at different bias voltages.\\
\textbf{Supplementary Figure \ref{ref Rs spectra}}. Responsivity spectra of the reference device at different bias voltages.\\
\textbf{Supplementary Figure \ref{Polariton OPD noise}}. Noise spectral density of polariton OPD.\\
\textbf{Supplementary Figure \ref{comparison}}. Comparison of key metrics of the polariton OPD (965~nm) and a reference device.\\
\textbf{Supplementary Table \ref{tab:responsivity}}. Overview of reported narrowband NIR-OPDs.\\
\newpage

\newpage

\newpage

\newpage

\newpage
\textbf{Polariton characteristics of the devices:}
Here is the analysis for polariton devices with resonances at 965 nm and 990 nm, analogous to the polariton OPD with a resonance at 945 nm, as shown in Fig.~\ref{Device ref analysis}. For the device with a resonance at 965 nm, the fitting results show a Rabi splitting of 0.55 eV and 0.32 eV, with a detuning value of $\Delta = E_{c} - E_{IEICO-4F} = 0.33$ eV. For the device resonant at 990 nm, the fitting yields a Rabi splitting of 0.52 eV and 0.34 eV, with a detuning of $\Delta = E_{c} - E_{IEICO-4F} = 0.26$ eV. The devices were tuned by adjusting the active layer thickness to approximately 99 nm and 110 nm for resonances at 965 nm and 990 nm, respectively. 

\newpage

\newpage

\newpage

\newpage

\newpage

\newpage

\newpage
\begin{table}[h!]
\centering
\begin{tabular}{|c|c|c|c|}
\hline
\textbf{Wavelength (nm)} & \textbf{Responsivity (A/W)} & \textbf{Active layer thickness (nm)} & \textbf{Reference} \\ \hline
940 & 0.24 &  2000 & 12 \\ \hline
945 & \textbf{0.23} & \textbf{92} & \textbf{This work} \\ \hline
950 & 0.06 & 1500 &  10 \\ \hline
950 & 0.05 & 220* &19\\ \hline
950 & 0.137 & $50^{\#}$ & 15 \\ \hline
960 & 0.14 & 440 &16\\ \hline
965 & \textbf{0.23}& \textbf{99} &\textbf{This work}  \\ \hline
980 & 0.14 &  500 & 13 \\ \hline
987 & 0.095 & $\sim$180 & 18 \\ \hline
990 & \textbf{0.11} & \textbf{110} &\textbf{This work} \\ \hline
990 & 0.0185 & - & 20 \\ \hline
\end{tabular}
\caption{Overview of reported narrowband NIR-OPDs to date with peak wavelength ranging from 940 to 990~nm at 0~V}
\label{tab:responsivity}
\end{table}

\noindent\textit{
\begin{itemize}
    \item[*] \textnormal{The overall device thickness was tuned using a spacer layer with a thickness of 220 nm.}
    \item[$^{\#}$] \textnormal{The total thickness of the device is 220 nm.}
\end{itemize}
}

\end{document}